\documentclass[sn-mathphys,Numbered]{sn-jnl}%

\usepackage{graphicx}%
\usepackage{multirow}%
\usepackage{comment}
\usepackage{amsmath,amssymb,amsfonts}%
\usepackage{amsthm}%
\usepackage{mathrsfs}%
\usepackage[title]{appendix}%
\usepackage{xcolor}%
\usepackage{textcomp}%
\usepackage{manyfoot}%
\usepackage{subfig}
\usepackage{graphicx}
\usepackage{booktabs}%
\usepackage{algorithm}%
\usepackage{algorithmicx}%
\usepackage{algpseudocode}%
\usepackage{dsfont}
\usepackage{listings}%

\begin{document}
\title{Understanding gravitationally induced decoherence parameters in neutrino oscillations using a microscopic quantum mechanical model}

\author*[1]{\fnm{Alba} \sur{Domi}}\email{alba.domi@fau.de}
\equalcont{These authors contributed equally to this work.}
\author*[1]{\fnm{Thomas} \sur{Eberl}}\email{thomas.eberl@fau.de}
\equalcont{These authors contributed equally to this work.}

\author*[1,2]{\fnm{Max Joseph} \sur{Fahn}}\email{max.j.fahn@fau.de}
\equalcont{These authors contributed equally to this work.}

\author*[1,2]{\fnm{Kristina} \sur{Giesel}}\email{kristina.giesel@fau.de}
\equalcont{These authors contributed equally to this work.}

\author*[1]{\fnm{Lukas} \sur{Hennig}}\email{lukas.hennig@fau.de}
\equalcont{These authors contributed equally to this work.}

\author*[1]{\fnm{Ulrich} \sur{Katz}}\email{uli.katz@fau.de}
\equalcont{These authors contributed equally to this work.}

\author*[1,2]{\fnm{Roman} \sur{Kemper}}\email{roman.kemper@fau.de}
\equalcont{These authors contributed equally to this work.}

\author*[1,2]{\fnm{Michael} \sur{Kobler}}\email{michael.kobler@fau.de}
\equalcont{These authors contributed equally to this work.}

\affil[1]{\orgdiv{Erlangen Centre for Astroparticle Physics}, \orgname{Friedrich-Alexander-Universität Erlangen-Nürnberg, Department of Physics}, \orgaddress{\street{ Nikolaus-Fiebiger-Straße 2}, \city{Erlangen}, \postcode{91058}, \country{Germany}}}

\affil[2]{\orgdiv{Friedrich-Alexander-Universität}, \orgname{Department of Physics, Theoretical Physics III, Institute for Quantum Gravity}, \orgaddress{\street{Staudtstr. 7}, \city{Erlangen}, \postcode{91058},  \country{Germany}}}

\abstract{In this work, a microscopic quantum mechanical model for gravitationally induced decoherence introduced by Blencowe and Xu is investigated in the context of neutrino oscillations. The focus is on the comparison with existing phenomenological models and the physical interpretation of the decoherence parameters in such models. The results show that for neutrino oscillations in vacuum gravitationally induced decoherence can be matched with phenomenological models with decoherence parameters of the form $\Gamma_{ij}\sim \Delta m^4_{ij}E^{-2}$. When matter effects are included, the decoherence parameters exhibit a dependence on the varying matter density across the Earth layers. This behavior can be explained by the nature of the coupling between neutrinos and the gravitational wave environment, as suggested by linearised gravity. On a theoretical level, these different models can be characterised by a different choice of Lindblad operators, with the model with decoherence parameters that do not include matter effects being less suitable from the point of view of linearised gravity. Consequently, in the case of neutrino oscillations in matter, the microscopic model does not agree with many existing phenomenological models that assume constant decoherence parameters in matter. Nonetheless, we identify the KamLAND experimental setup as particularly well-suited to establish the first experimental constraints on the model parameters, namely the neutrino coupling to the gravitational wave environment and its temperature, based on a prior analysis using the phenomenological model.}


\maketitle
\newpage
\section{Introduction}\label{sec:Intro}
The search for quantum decoherence (QD) effects in connection with neutrino oscillations is a topic that has gained increasing interest in various research communities in recent years \cite{Gago:2000qc,Benatti:2000ph,Klapdor-Kleingrothaus:2000kdx,Lisi:2000zt,Benatti:2001fa,Morgan:2004vv,Anchordoqui:2005gj,Mavromatos:2006yy,Mavromatos:2007hv,Fogli:2007tx,Farzan:2008zv,Oniga:2015lro,BalieiroGomes:2016ykp,Oliveira:2016asf,Carpio:2017nui,Coelho:2017byq,Coelho:2017zes,BalieiroGomes:2018gtd,Carpio:2018gum,Carrasco:2018sca,Coloma:2018idr,Gomes:2020muc,Lessing:2023uxb,icecube_qd,minos_t2k,Buoninfante:2020iyr,Lagouvardos:2020laf,Ohlsson:2020gxx,Stuttard:2020qfv,Stuttard:2021uyw,Banerjee:2022slh,DeRomeri:2023dht}. Neutrino experiments have searched for indications of QD effects in current data via a phenomenological approach \cite{Lessing:2023uxb,Coloma:2018idr,Gomes:2020muc,icecube_qd, minos_t2k,DeRomeri:2023dht} and several works have analysed the sensitivity of future detectors \cite{Carpio:2018gum,Barenboim:2024wdn}. However, the connection between such phenomenological models and the underlying microscopic physics is not always immediate. It is interesting to understand, in terms of theoretical models in the framework of open quantum systems \cite{Breuer:2007juk}, how the phenomenological models used for neutrino oscillations can be derived from underlying microscopic physics. 
~\\
~\\
Existing phenomenological models often start from the Lindblad equation and then parameterise the dissipator involved by selecting a finite number of parameters. In the most general, three-neutrino case, the number of free parameters is $45$, but this number is usually reduced by requiring physically meaningful additional assumptions \cite{Oliveira:2016asf,BalieiroGomes:2016ykp,BalieiroGomes:2018gtd} such as unitarity or conservation of energy of the neutrino subsystem\footnote{Conservation of energy in the overall system of neutrino and environment is of course always fulfilled}, see for instance \cite{DeRomeri:2023dht} for a recent brief introduction to these models. These assumptions reduce the number of free parameters characterising the dissipator in the current phenomenological models to less than five, see for instance \cite{DeRomeri:2023dht,Lessing:2023uxb}.
~\\
~\\
From a theoretical perspective, the interest lies in the underlying microscopic model, from which the Lindblad equation can be derived under various assumptions about the open quantum system. A crucial choice in such microscopic models is the environment as well as the specific coupling of the environment and the system under consideration encoded in the interaction Hamiltonian. At the level of the Lindblad equation, the choice of coupling plays a role in two ways: firstly, the choice of coupling between the system and environment can be linked to some choice of Lindblad operators, see also \cite{Xu:2020pzr} in the context of neutrino oscillations. Secondly, the environmental operators involved in the interaction Hamiltonian determine the specific form of the environmental correlation functions when the environmental degrees of freedom are traced out, which in turn determine the coefficients of the Lindblad operators involved in the dissipator.
\\ \\
In the framework of existing phenomenological models, which usually start at the level of the Lindblad equation, the dissipator for a three-neutrino scenario is often parameterised by the choice of eight dissipator operators, which is then transferred to the choice of a finite number of decoherence parameters, usually less than five. These decoherence parameters thus contain all the information about the choice of Lindblad operators and the form of the correlation functions of the environment. Starting from a microscopic model, the appropriate form and number of decoherence parameters can be calculated, and one has sufficient control over their physical interpretation and the assumptions that go into this calculation. However, linking a given set of decoherence parameters to an underlying microscopic model is more difficult, since the degrees of freedom of the environment have already been traced out and reconstructing the coupling between the system and the environment in this direction is less straightforward and unambiguous.
\\ \\
Moreover, a large class of phenomenological models assumes that the contribution in the dissipator, which is responsible for decoherence and usually leads to a damping of the probabilities for neutrino oscillations, depends on a positive or negative power of the neutrino energy  \cite{Lisi:2000zt,Fogli:2007tx,Coloma:2018idr,Carrasco:2018sca,Gomes:2020muc,Stuttard:2020qfv,Lessing:2023uxb}. Different choices of the power define different models that generally yield other modifications of the oscillation probabilities.
\\
From a quantum gravity perspective, the underlying microscopic models are of interest because they allow the formulation of models in which gravitationally induced decoherence can be investigated \cite{Klapdor-Kleingrothaus:2000kdx,Mavromatos:2006yy,Mavromatos:2007hv,Stuttard:2020qfv,Stuttard:2021uyw,Banerjee:2022slh}, see also \cite{Bassi:2017szd,Anastopoulos:2021jdz} for reviews on gravitational decoherence. In the context of general relativity, a suitable starting point for the formulation of such microscopic models is linearised gravity coupled to a matter system, and in the framework of open quantum systems, the corresponding master equation can be derived in quantum field theory  \cite{Anastopoulos:2013zya,Fahn:2022zql,Lagouvardos:2020laf,Oniga:2015lro,Blencowe:2012mp}. Since all these models contain an infinite number of degrees of freedom, the final master equations are rather complicated and difficult to handle for the case of neutrino oscillations. A possible solution is to consider the 1-particle sector of the field theory model and use this as the underlying microscopic model in a quantum mechanical setting in which the existing phenomenological models work, see for instance \cite{Fahn:2024fgc} for an analysis of the 1-particle sector of the quantum field theory model from \cite{Fahn:2022zql} and also \cite{Anastopoulos:2013zya,Lagouvardos:2020laf}. In \cite{Fahn:2024fgc} the projection of the field theoretical model with an infinite number of degrees of freedom onto its 1-particle sector is considered, which in turn can be formulated as a quantum mechanical model involving only finitely many degrees of freedom. 
~\\ ~\\
As a first step towards bridging the gap between the underlying microscopic models and the existing phenomenological ones, this work considers a quantum mechanical toy model for gravitationally induced decoherence introduced by Blencowe and Xu \cite{Xu:2020lhc} and slightly generalised it in order to apply it in the context of neutrino oscillations. The model in \cite{Xu:2020lhc} is inspired by the models in \cite{Anastopoulos:2013zya,Fahn:2022zql,Lagouvardos:2020laf,Oniga:2015lro,Blencowe:2012mp} as far as the choice of the environment and its coupling to a given matter system is concerned. To the authors knowledge, the model in \cite{Xu:2020lhc}, that was already briefly discussed in \cite{Blencowe:2012mp}, has been investigated in the context of neutrino oscillations so far only in \cite{DEsposito:2023psn} where they conclude that this model leads to no decoherence effect. However, to our understanding, as it is discussed below, with the assumption we use in this work there is a non-vanishing decoherence effect. 
~\\ ~\\
In this work, the only free parameters that remain in the final Lindblad equation are the coupling parameter of the neutrino to the environment and a temperature parameter that characterises the environment of the thermal gravitational waves. In this way, it is possible to obtain a physical interpretation of the free parameters in the existing phenomenological models. Moreover, the gravitationally induced decoherence model presented in this work favours a certain power of the neutrino energy dependence in the dissipator. In the phenomenological model, this energy dependence is postulated, whereas, in this work it directly follows from the choice of the underlying microscopic model. In addition, as our results show, there is a clear physical interpretation of the decoherence parameters present in a subclass of phenomenological models. Interestingly, as we will explain below, the existing bounds for the decoherence parameters in the phenomenological models in \cite{DeRomeri:2023dht} for the KamLAND experiment can be used to constrain the free parameters of the microscopic model considered in this work.
~\\ ~\\
In the most general case, and also in the model presented here, the Lindblad equation contains a so-called Lamb shift contribution, which is caused by the interaction with the gravitational environment. This contribution modifies the unitary evolution of the effective dynamics of the neutrinos and does neither lead to dissipation nor decoherence. However, it has the effect that the energy eigenvalues of the neutrinos are modified by a shift, which in the model considered here depends on a cutoff frequency. The latter is introduced as a regulator for some of the integrals involved in the calculation of the model's environmental correlation functions. As an energy shift depending on some regulator is rather inconvenient, we will show that along the lines of the Caldeira-Leggett model of quantum Brownian motion \cite{Caldeira:1981rx} the Lamb shift contribution can be eliminated by a suitable counter-term. Such a Lamb shift contribution is also present in the field theoretical models in \cite{Anastopoulos:2013zya,Fahn:2022zql,Lagouvardos:2020laf,Oniga:2015lro,Blencowe:2012mp} where a similar renormalisation procedure is applied. In many existing phenomenological models such a Lamb shift contribution is often just neglected and no counter term is considered. Our analysis shows that such a strategy is justified for the model considered here, but in general requires a detailed investigation for each individual model considered, as in general the renormalisation of the Lamb shift term can provide additional contributions. Our results further show that the interpretation of the Lamb shift contribution discussed in \cite{Benatti:2001fa} in the form of massless neutrino oscillations is somewhat problematic, as we understand it, when carried over to the model here or the field-theoretic models in \cite{Anastopoulos:2013zya,Fahn:2022zql,Lagouvardos:2020laf,Oniga:2015lro,Blencowe:2012mp}.
~\\ ~\\
The article is structured as follows: after the introduction in Sec.~\ref{sec:Intro} we briefly introduce in Sec.~\ref{sec:MicrosModelGen} the model from \cite{Xu:2020lhc} and its slight generalisation which is needed for the further analysis. Here we discuss the model for a generic choice of matter system with only very mild assumptions that are also consistent with those used in the existing phenomenological models. Here we skip details of the derivation that can partly be found in the appendix but discuss what kind of assumptions enter into the model and how these are motivated. The application of the model to a three neutrino-scenario is presented in Sec.~\ref{sec:MicrosModelNeutrino}. It provides the solution of the effective neutrino dynamics under the influence of the gravitational environment that is taken as starting point in Sec.~\ref{sec:Results} to compute the probabilities for neutrino oscillations that follow from the model considered in this work. The main focus in this section lies on the comparison of the model considered in this work and the existing phenomenological models. As will be shown they can be understood as two models with different couplings to the gravitational environment that agree for the special case of neutrino oscillations in vacuum. In the matter case, deviations in the oscillation probabilities appear, which are discussed and quantified. The first bounds on the model parameters are derived from a previous phenomenological analysis performed on KamLAND data \cite{DeRomeri:2023dht}.
Finally, in Sec.~\ref{sec:Concl} the main results of this work are summarised and an outlook on possible future work in this direction is presented.

\section{Decoherence model}\label{sec:Model}
The model investigated in this work is inspired from the field theory models for gravitationally induced decoherence \cite{Anastopoulos:2013zya,Fahn:2022zql,Lagouvardos:2020laf,Oniga:2015lro,Blencowe:2012mp} that all consider linearised gravity as the environment to which a given matter system is coupled. Because general relativity as well as generic matter systems involve gauge symmetries, some work is necessary in order to get the corresponding physical Hamiltonian of the total system that usually is the starting point of the decoherence model. This has been implemented in \cite{Anastopoulos:2013zya,Fahn:2022zql,Lagouvardos:2020laf,Oniga:2015lro,Blencowe:2012mp} either by gauge fixing  \cite{Anastopoulos:2013zya,Lagouvardos:2020laf,Oniga:2015lro,Blencowe:2012mp} or by the construction of gauge invariant observables \cite{Fahn:2022zql} by means of choosing a suitable dynamical reference system. The classical total Hamiltonian $H_{\rm tot}$ in all these models has the form
\begin{equation*}
H_{\rm tot} = H_{S} + H_{\cal E} + H_{\rm int},  
\end{equation*}
where $ H_{S}$ encodes the dynamics of the system usually chosen to be some matter, $H_{\cal E}$ is the Hamiltonian for the environment, here linearised gravity, and $H_{\rm int}$ describes their interaction. Once a frame of reference has been chosen, the form of each Hamiltonian can be derived from the underlying action, and in particular the form of the interaction Hamiltonian $H_{\rm int}$ is determined by the way matter and (linearised) gravity are coupled, namely via the energy-momentum tensor of matter and the metric. Thus, on the one hand it is an advantage to know the underlying field theory model because the microscopic Hamiltonian that enters any decoherence models can rather be derived than needing to be chosen, which results in decoherence models with less ambiguities. On the other hand as the results in \cite{Anastopoulos:2013zya,Fahn:2022zql,Lagouvardos:2020laf,Oniga:2015lro,Blencowe:2012mp} illustrate, the final form of the master equation which encodes the dynamics of the system's density matrix when the environmental degrees of freedom have been traced out, is very complicated and hence technically challenging. Therefore, as a first step in this work we will consider the quantum mechanical toy model for gravitationally induced decoherence introduced in the seminal work of Xu and Blencowe in \cite{Xu:2020lhc}. This model is strongly inspired by the field theory models and hence mimics that usual gravitational coupling in the quantum mechanical toy model. To the knowledge of the authors although this model exists in the literature it has only been applied to investigate gravitationally induced decoherence in the context of neutrino oscillations in \cite{DEsposito:2023psn} where the authors however conclude that the model will lead to no decoherence effect if they apply the equal-energy condition motivated in their work from the wave packet approach.  We will consider the slightly generalised model of \cite{Xu:2020lhc} such that we can also apply it to neutrino oscillations and present the derivation of the corresponding master equation, which was not included in the presentation in \cite{Xu:2020lhc} in detail to show that from our results non-vanishing decoherence effects are possible for this model.

\subsection{Microscopic model for generic time independent system's Hamiltonian}
\label{sec:MicrosModelGen}
\begin{figure}[h]
\centering
\subfloat{\label{main:aX}\includegraphics[scale=.8]{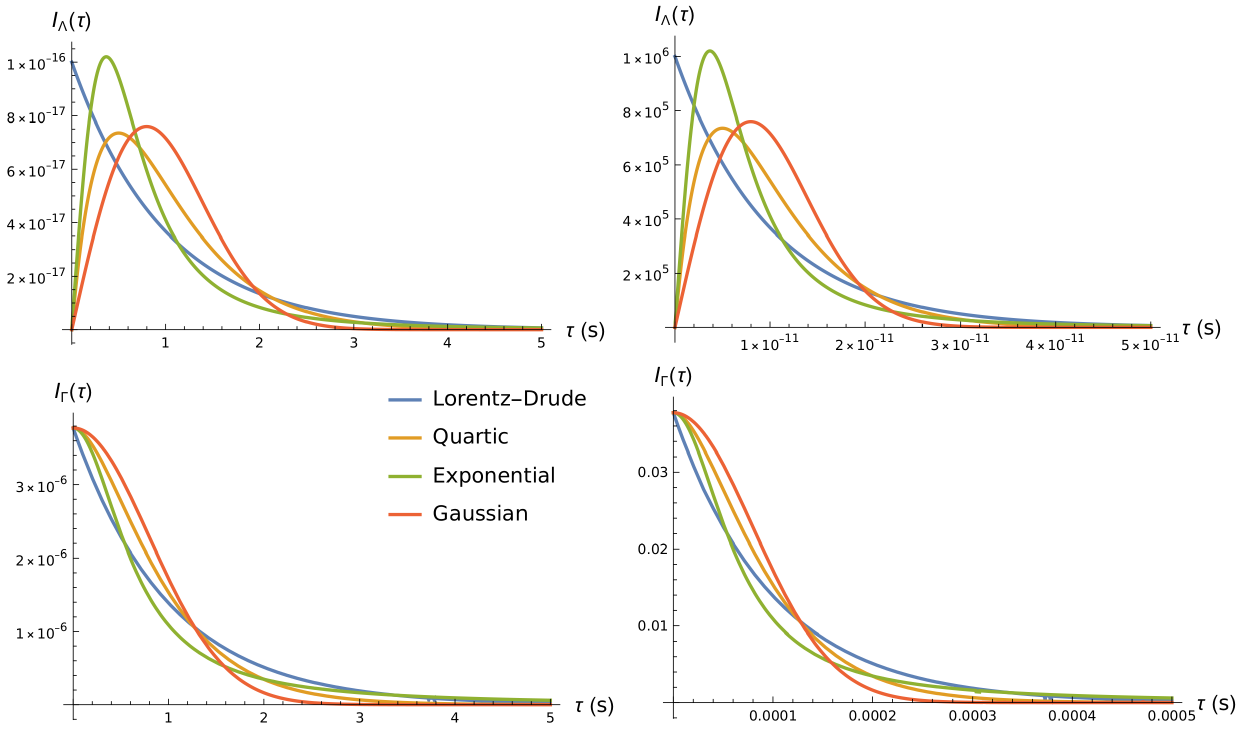}}
\caption{Integrands $I_\Lambda(\tau)$ (first row) and $I_\Gamma(\tau)$ (second row) defined in ~\eqref{eq:DefintLam} and ~\eqref{eq:DefintGam} for the different cutoff functions and with parameter values $\eta = 10^{-8}\,\mathrm{s}$, $T=0.9\,\mathrm{K}$ and $\Omega=1\,\mathrm{Hz}$ (first column), $\Omega=100 \,\mathrm{GHz}$ (second column, first row), and $\Omega=10\,\mathrm{kHz}$ (second column, second row). }
\label{fig:integrandsMarkAp}
\end{figure}

In \cite{Xu:2020lhc} the Hamiltonian of the system is chosen to be a harmonic oscillator. The effective dynamics of the system is derived by choosing the bath, thus the N harmonic oscillators in the environment to be in a thermal state, which is then expressed in terms of the coherent state basis, while for the density matrix of the system the number basis is chosen.
~\\
Here, we want to consider the model in a more general context to be able to apply it later to neutrino oscillations. Therefore, we will leave the choice of the system's Hamiltonian generic in this section and only specify to neutrinos later. The only assumption we make for $\hat{H}_{S}$ is that it is time-independent. 
To motivate the remaining contributions of the total Hamiltonian of the toy model, we consider that in general relativity the metric couples to the energy-momentum-tensor of the matter field which itself depends on the metric. From a physical point of view, we expect that the gravitational influence on neutrino oscillations, as considered in this article, represents only a small perturbation of the free dynamics of the neutrino and thus the coupling between matter and gravity is assumed to be weak. Therefore, we can focus on the linearised formulation of gravity in which we consider flat Minowski spacetime as the background and consider perturbations around it. From the perturbed action we realise that for the linearised case the linear perturbation of the metric couple to the energy-momentum tensor of the matter field evaluated on the flat Minkowski background, see for instance \cite{Fahn:2022zql}. The physical gravitational degrees of freedom are fully described by gravitational waves that for the two existing polarisations describe for each Fourier mode a harmonic oscillator. This is the reason why the choice was made in \cite{Xu:2020lhc} to model the gravitational environment in the toy model as a collection of $N$ uncoupled harmonic oscillators. If they were coupled, this would also include higher orders than the linear one in the corresponding field theoretical model, which, however, are expected to be negligibly small due to the higher order in the weak coupling constant. Next the coupling between the N harmonic oscillators and the systems via the position operators of the environment is also motivated from linearised gravity. There, in the Hamiltonian formulation \cite{Deser:1960zzc}, the metric components become the configuration variables and it is precisely their linear perturbations that couple to the energy momentum tensor in the linearised theory. To mimic such a kind of coupling in the quantum mechanical toy model,  the Hamiltonian of the matter system $\hat{H}_S$, as a substitute for the energy-momentum tensor, is coupled to the position operators of the N harmonic oscillators, where the latter are a substitute for operators corresponding to the metric perturbations. The total Hamiltonian then splits into the individual contributions which are quantised in a quantum mechanical context yielding
\begin{equation}
\label{eq:MicroHam}
    \hat{H} =  \hat{H}_S + \hat{H}_{\cal E} + \hat{H}_{\rm int}
   = \underbrace{\hat{H}_{S}^{(0)} + \hat{H}_{S}^{(C)}}_{\hat{H}_{S}} + \underbrace{\frac{1}{2}\sum_{i=1}^N \left[ \hat{p}_i^2 + \omega_i^2 \hat{q}_i^2 \right]}_{\hat{H}_{\cal E}} - \underbrace{\hat{H}_{S} \otimes \sum_{i=1}^N g_i \hat{q}_i}_{\hat{H}_{\rm int}}\,,
\end{equation}
where here the specific form of $\hat{H}_{\cal E}$ and $\hat{H}_{\rm int}$ are inspired from the field theory model as discussed above and $\hat{H}_{S}^{(C)}$ denotes a counter term of the form $\hat{H}_{S}^{(C)}= \sum_{i=1}^N \frac{\hbar g_i^2}{2\omega_i^2} (\hat{H}_{S}^{(0)})^2$. This counter term is needed as it will later remove the unphysical contribution of the Lamb shift contribution. It is included analogously to the treatment of the Caldeira-Leggett model \cite{Caldeira:1981rx}, see for instance \cite{Xu:2020lhc} and can be understood as a tiny, due to $g_i^2$, frequency dependent correction to the unitary evolution of the non-renormalised and thus the bare system's Hamiltonian $\hat{H}_{S}^{(0)}$. The inclusion of a counter term and the necessity of renormalisation also arise in  similar field theoretical models, see for instance \cite{Fahn:2024fgc}. In that work the impact of  renormalisation is analysed in the context of a field theoretical model and the differences to the renormalisation of the quantum mechanical toy model in the present work are discussed in detail.
\\
In later applications in the Sec.~\ref{sec:MicrosModelNeutrino}, the system is chosen as neutrinos. The coupling constant $g_i$, which has dimension of inverse length, can in principle be different for each oscillator.
Position and momentum operators of the oscillators in the environment fulfil the usual commutation relations:
\begin{equation}
    \left[\hat{q}_i,\hat{p}_j\right] = i\hbar \delta_{ij}\,.
\end{equation}
The entire Hilbert space is a tensor product of the system Hilbert space $\mathcal{H} = \mathcal{H}_S \otimes \mathcal{H}_{\cal E}$ where $\hat{H}_S$ and $\hat{H}_{\cal E}$ act trivially on ${\cal H}_{\cal E}$ and ${\cal H}_S$, respectively. Assuming that the interaction is small (i.e. $g_i$ small) compared to the evolution in the absence of coupling with the environment, a time-convolutionless (TCL) master equation truncated to second order in the coupling $g_i$ (see e.g. \cite{Breuer:2007juk,Fahn:2022zql}) provides a good approximation to the effective dynamics of the system, which is obtained after the degrees of freedom of the environment have been worked out. 
In the model discussed here, this master equation then assumes a simple form, since the interaction Hamiltonian $H_{\rm int}$ contains a time-independent system Hamiltonian $\hat{H}_S$:
\begin{align}
\label{eq:MasterEqn}
    \frac{d}{dt}\hat{\rho}_S(t) =& -\frac{i}{\hbar}\left[ \hat{H}_S, \hat{\rho}_S(t)\right] - \frac{1}{\hbar^2} \int_0^{t-t_0} d\tau \; \text{Tr}_{\cal E}\left( \left[ \hat{H}_{\rm int}, \left[ \hat{H}_{\rm int}(-\tau), \hat{\rho}_S(t) \otimes \hat{\rho}_{\cal E} \right] \right] \right) \\ \nonumber
    =&-\frac{i}{\hbar} \left[ \hat{H}_S, \hat{\rho}_S(t)\right] + \frac{i \Lambda(t-t_0)}{\hbar^2} \left[  (\hat{H}_S^{(0)})^2, \hat{\rho}_S(t)\right]  +\frac{\Gamma(t-t_0)}{\hbar^2} \left( \hat{H}_S^{(0)} \hat{\rho}_S(t) \hat{H}_S^{(0)} - \frac{1}{2} \left\{ (\hat{H}_S^{(0)})^2, \hat{\rho}_S(t) \right\} \right)\,,
\end{align}
where $[.,.]$ and $\{.,.\}$ denote a commutator and an anticommutator, respectively. $\text{Tr}_{\cal E}(.)$ denotes the partial trace over the degrees of freedom of the environment that can be calculated once a state has been chosen for the environment, where we choose a Gibbs state i.e. $\hat{\rho}_{\cal E} = \frac{1}{Z} e^{-\beta \hat{H}_{\cal E}}$ with partition function $Z= \text{Tr}_{\cal E} \left( e^{-\beta \hat{H}_{\cal E}}\right)$, where $\beta = \frac{1}{k_B T}$ with the Boltzmann constant $k_B$ and similar to \cite{Anastopoulos:2013zya} we denote the involved 'temperature' parameter of the environment by $T$, which characterises the bath of the oscillators in the environment that mimic the thermal gravitational wave background in this toy model\footnote{Note that this parameter is denoted as $\Theta$ in the models in \cite{Anastopoulos:2013zya,Anastopoulos:2021jdz,Fahn:2022zql}.}. The density matrix of the matter system evaluated at temporal coordinate $t$ is denoted by $\hat{\rho}_S(t)$ and $\hat{H}_{\rm int}(\tau)$ is the interaction Hamiltonian operator in the interaction picture evaluated at time $-\tau$, i.e. $\hat{H}_{\rm int}(\tau) : = e^{\frac{i}{\hbar}(\hat{H}_S+\hat{H}_{\cal E})\tau} \hat{H}_{\rm int} e^{-\frac{i}{\hbar}(\hat{H}_S+\hat{H}_{\cal E})\tau}$. 
 To derive the master equation in \ref{eq:MasterEqn}, we have assumed factorising initial conditions, i.e. $\hat{\rho}(t_0) = \hat{\rho}_S(t_0) \otimes \hat{\rho}_{\cal E}$. As the master equation is truncated after second order in the couplings $g_i$, all terms of $\hat{H}_S^{(C)}$ vanish in the second and third term. A detailed discussion of the derivation can be found, for example, in \cite{Fahn:2022zql}. 
~\\ ~\\
The first term of the master equation is the standard unitary evolution of the matter system itself. The second term, usually referred to as the Lamb shift contribution, leads to a renormalisation of the energy levels of the matter systems due to the presence of the environment, and the third term is the dissipator present in open quantum systems. An analogue contribution to the Lamb shift, which results here directly from the derivation of the master equation, is also present in the field theoretical model. In addition in the field theory model a further gravitationally induced self-interaction term for the matter system is present, such a term is not involved in the quantum mechanically toy model because on the one hand it is strongly related to the gauge symmetries in general relativity and the construction of gauge invariant quantities, see for instance \cite{Fahn:2022zql,Fahn:2024-photons} and on the other hand whether it is present in the 1-particle projection also depends on the normal ordering choosing in the field theory model, see the discussion in \cite{Fahn:2022zql,Fahn:2024fgc}. In the field theoretical case, a renormalisation must normally be carried out for this contribution. The dissipator resembles the effective interaction of the gravitational wave environment with the matter system. Both contributions would be absent if one treats the neutrinos as a closed quantum system.
The latter two contributions involve two functions $\Lambda(t-t_0)$ and $\Gamma(t-t_0)$, respectively, both of which depend on a time interval $t-t_0$, where $t$ is the time at which the master equation is evaluated and $t_0$ is the initial time. These two functions are explicitly given by
\begin{align}\label{eq:lambdadef}
    \Lambda(t-t_0) &= \int_0^{t-t_0} d\tau \int_0^\infty d\omega \;\sin(\omega\tau) J(\omega) \\
    \Gamma(t-t_0) &= 2 \int_0^{t-t_0} d\tau \int_0^\infty d\omega \;\cos(\omega\tau) \coth\left(\frac{\hbar\omega}{2 k_B T}\right) J(\omega) \label{eq:gammadef}\,,
\end{align}
where $J(\omega)$ is the spectral density that characterises the strength with which different frequencies in the environment contribute to the interaction with the matter system. From the model, it is given as
\begin{equation}\label{eq:jsumdef}
    J(\omega) = \frac{1}{2} \sum_{i=1}^N g_i^2 \frac{\hbar^2}{\omega_i} \delta(\omega-\omega_i)\,,
\end{equation}
where $\delta(.)$ denotes the Dirac delta function. Given that not all the oscillators in the environment are neither known in detail, nor of interest, one often approximates $J(\omega)$ by a smooth function in $\omega$, see e.g. \cite{Breuer:2007juk}. The usual requirements for this function are that it is linear in $\omega$ for small $\omega$ and that it tends to zero for large $\omega$. Note that such a spectral density is also chosen, for example, in the Caldeira-Leggett model for quantum Brownian motion, and the linear dependence is crucial in this case to obtain the friction term present in that model after renormalisation. Such spectral density with such a linear behavior are usually called Ohmic spectral densities. Many models used in the existing literature \cite{Breuer:2007juk,weiss2012quantum,Xu:2020lhc} use an Ohmic spectral density and differ only by the chosen cutoff function, see for instance also \cite{Grabert:1984DHO} for an application of the Drude regularisation. Note that a different than linear behaviour for small $\omega$ would for the model considered here lead to IR divergences for  all $\omega^m$ with $m\leq 0$, $m\in\mathbb{Z}$ and partly also to a rather not physically reasonable scaling with the inverse temperature parameter $T$ in the decoherence term. The latter corresponds to a strong decoherence effect when the temperature is low that becomes infinite for $T=0$, that is when the Gibbs state corresponds to a vacuum state. For  $m\geq 2$ with $m\in\mathbb{N}$ there exist no decoherence effects in the model considered in this work.
~\\ ~\\
 A comparison to the field-theoretic model derived in \cite{Fahn:2022zql}, which motivates this toy model here, shows that linearity in $\omega$ for small $\omega$ is reasonable, while a cutoff for larger $\omega$ corresponds to the UV-divergencies in quantum field theory that also have to be renormalised. In appendix \ref{secA1} we discuss this comparison in more detail. 
~\\ ~\\
Since the choice of a particular spectral density is an assumption that goes into every model, we were interested in how the final result of the master equation depends on this choice. Therefore, we considered a few choices in this work that also include the most prominent ones used in the existing literature like the Lorentz-Drude and the exponential cutoff. These are shown below\footnote{There is a typo in the published version, as the definition of spectral density differs by a factor of two from the one used here.}:
\begin{align}
\text{Lorentz-Drude cutoff: } J(\omega) &=\frac{4}{\pi} \eta^2 \omega \frac{\Omega^2}{\Omega^2+\omega^2}\\
\text{Quartic cutoff: } J(\omega) &=\frac{4}{\pi} \eta^2 \omega \frac{(2\Omega)^4}{((2\Omega)^2+\omega^2)^2}\\
\text{Exponential cutoff: } J(\omega) &=\frac{4}{\pi} \eta^2 \omega e^{-\frac{2\omega}{\pi\Omega}}\\
\text{Gaussian cutoff: } J(\omega) &=\frac{4}{\pi} \eta^2 \omega e^{-\frac{\omega^2}{\pi\Omega^2}}\,.
\end{align}
In these functions, $\eta^2$ is a free parameter which we will discuss further below and $\Omega$ is a cutoff frequency used as regulator or the otherwise divergent integrals. Evaluating the $\int d\omega$ integral in  \eqref{eq:lambdadef} and  \eqref{eq:gammadef} yields the integrands $I_\Lambda(\tau)$ and $I_\Gamma(\tau)$ defined as
\begin{align}\label{eq:DefintLam}
    \Lambda(t-t_0) &=  \int_0^{t-t_0} d\tau \; I_\Lambda(\tau)\\
    \Gamma(t-t_0) &= 2 \int_0^{t-t_0} d\tau \; I_\Gamma(\tau)\,.\label{eq:DefintGam}
\end{align}
A plot of these functions is shown in Fig.~\ref{fig:integrandsMarkAp}.

\noindent As discussed below in Sec.~\ref{sec:ResRenorm}, for thermal gravitational waves a reasonable cutoff frequency is $\Omega=100\,\mathrm{GHz}$. Hence, in the case of $I_\Lambda(\tau)$, the integrand decays rapidly on timescales of $\tau=10^{-10}\,\mathrm{s}$ which is much smaller than the timescale on which the core system varies, that is determined for the neutrinos investigated in this work by $\left(\frac{1}{\hbar} \frac{\Delta m^2}{2 E}\right)^{-1}$, which\footnote{As the dominant contribution to the neutrino evolution comes from the part $e^{\frac{i}{\hbar} \frac{\Delta m^2}{2 E} t}$ and the matter effects have a similar order of magnitude.} is around the order of magnitude of $1\,\mathrm{s}$. Furthermore, the plots for $I_\Gamma(\tau)$ suggest that $\Gamma(t-t_0)$ is independent of $\Omega$ for $t-t_0$ sufficiently large enough, as the two axes scale inversely to each other with the same ratio. Already for $\Omega=10\,\mathrm{kHz}$, where $I_\Gamma(\tau)$ is numerically still well computable, the timescale on which $I_\Gamma(\tau)$ decays is much lower than the one on which the system state varies. This means that the environment rapidly "forgets" about the history of the system and thus the Markovian approximation for a memoryless process is justified. Therefore, in both cases the error committed when shifting the initial time $t_0\rightarrow -\infty$ and hence the upper integration limit of the $\tau$ integration to $+\infty$ is negligible. Given that, we perform the so-called second Markov approximation for the further analysis, which then leads to time-independent results for $\Lambda$ and $\Gamma$: 
\begin{align}
    \Lambda &:= \Lambda(\infty) = 2\Omega \eta^2 \\
    \Gamma &:= \Gamma(\infty) = \frac{8 \eta^2k_B T}{\hbar}\,.
\end{align}
As expected, $\Gamma$ is independent of $\Omega$.
When introducing any of the above named spectral densities, the corresponding counter term is the same for all choices of the cutoff in the spectral density and given by:
\begin{equation}
\label{eq:CounterTerm}
   \hat{H}_{S}^{(C)} = \frac{1}{\hbar} \int_0^\infty d\omega \frac{J(\omega)}{\omega} (\hat{H}_{S}^{(0)})^2 = \frac{2\Omega\eta^2}{\hbar}  (\hat{H}_{S}^{(0)})^2\,,
\end{equation}
which precisely cancels the Lamb-shift contribution in the second order TCL master equation in Markovian limit yielding:
\begin{align}\label{eq:MasterEqMarkovLim}
    \frac{d}{dt}\hat{\rho}_S(t) =&-\frac{i}{\hbar} \left[ \hat{H}_S^{(0)}, \hat{\rho}_S(t)\right]  +\frac{8\eta^2}{\hbar^2} \frac{k_B T}{\hbar} \left( \hat{H}_S^{(0)} \hat{\rho}_S(t) \hat{H}_S^{(0)} -\frac{1}{2} \left\{ (\hat{H}_S^{(0)})^2, \hat{\rho}_S(t) \right\} \right)\,.
\end{align}
This master equation includes still two free parameters,  $\eta$ which is related to the coupling parameter between the system and the environment and the environmental temperature parameter $T$  that enters via $\beta=\frac{1}{Tk_B}$ in the Gibbs state, where $k_B$ denotes the Boltzmann constant. 
~\\ ~\\
In order to compare the master \eqref{eq:MasterEqMarkovLim} better to the existing phenomenological models later, we note that \eqref{eq:MasterEqMarkovLim} can be written in Lindblad form
\begin{align}\label{eq:MasterLindblad}
    \frac{d}{dt}\hat{\rho}_S(t) =&-\frac{i}{\hbar} \left[ \hat{H}_S^{(0)}, \hat{\rho}_S(t)\right]  +\frac{8\eta^2}{\hbar^2} \frac{Tk_B}{\hbar} \left( \hat{L} \hat{\rho}_S(t) \hat{L}^\dagger -\frac{1}{2} \left\{ \hat{L}^\dagger\hat{L}, \hat{\rho}_S(t) \right\} \right)\,.
\end{align}
with the choice of the Lindblad operator $\hat{L}=\hat{H}_S^{(0)}$ using that $\hat{H}_S^{(0)}$ is self-adjoint, that is $\hat{L}^\dagger=\hat{L}$ and thus $\hat{L}^\dagger\hat{L}=\hat{L}^2$,  see for instance \cite{Gambini:2003pv} where also a Lindblad operator proportional to the Hamiltonian of the system is chosen for a decoherence model inspired by discrete quantum gravity. In many phenomenological models, the Lindblad equation is taken as the starting point, often omitting the second term containing the contributions of the Lamb shift as well as a counter term. Then the model is characterised by a selection of Lindblad operators $\hat{L}_i$, of which there can generally be more than one, usually chosen to be either linear with the position and/or momentum operators of the matter system such that they can be written linearly in annihilation and creation operators see for instance \cite{Breuer:2007juk} for the standard examples or \cite{Smirne:2022uwk} for a non-perturbative treatment of multi-time expectation values in open quantum systems for specific environments.  The advantage of starting with a microscopic model as in \eqref{eq:MicroHam} is that, for example, the functions $\Gamma(t-t_0)$ and $\Lambda(t_0)$ can be derived and calculated directly, resulting in a model with less ambiguities at the level of the Lindblad equation. Furthermore, the choice of the Lindblad operator, following the field theoretical models \cite{Anastopoulos:2013zya,Fahn:2022zql,Lagouvardos:2020laf,Oniga:2015lro,Blencowe:2012mp}, is directly linked to the property of how linearised gravity couples to matter, and therefore in this sense is also determined by the microscopic model. We discuss in the application to neutrino oscillations a comparison between the renormalised and non-renormalised model in Sec.~\ref{sec:ResRenorm}. Finally, it is worth noting that for the special case of zero temperature $T=0$, in which the Gibbs state is only the vacuum state and thus the gravitational environment is assumed to be in a vacuum state, no decoherence effects are present, since in the model presented here the dissipator is linear in $T$. Note that this is a property of the gravtiational environment and independent of the chosen system under consideration and will thus also apply to the case of neutrinos in next section.

\subsection{Application to neutrino oscillations}
\label{sec:MicrosModelNeutrino}
In this section, we evaluate and solve the master equation \eqref{eq:MasterEqMarkovLim} for neutrinos with three different flavors as the matter system. To adapt the model to neutrinos choose the following system's Hamiltonian operator
\begin{equation}\label{eq:vacmb}
    \hat{H}_S^{(0)} = \hat{H}_{vac} +  \hat{U}^\dagger \hat{H}_{mat} \hat{U},
\end{equation}
where the second term takes into account that the neutrinos propagates through the (different layers of the) Earth and the neutrino Hamiltonian in vacuum in the mass basis is given by
\begin{equation}\label{eq:hvaco}
    \hat{H}_{vac} = \begin{pmatrix}
        E_1 & 0 & 0 \\ 0 & E_2 & 0 \\ 0 & 0 & E_3
    \end{pmatrix} = E \mathds{1}_3 + \frac{c^4}{6E} \begin{pmatrix}
    -\Delta m_{21}^2 - \Delta m_{31}^2 & 0 & 0 \\ 0 & \Delta m_{21}^2 -\Delta m_{32}^2 & 0 \\ 0 & 0 & \Delta m_{31}^2 + \Delta m_{32}^2 \end{pmatrix}\,,
\end{equation}
where we used that  $E_i \approx E + \frac{ m_{i}^2 c^4}{2E} $ and that we can modify the Hamiltonian by a constant matrix such that the difference $(E_i-E_j)$ is not modified because in the final equation only powers of the energy difference will contribute. Here the mass differences squared are denoted by $\Delta m_{ij}^2 = m_i^2-m_j^2$, the mean neutrino energy by $E=\frac{1}{3}(E_1+E_2+E_3)$ and the PMNS matrix by $\hat{U}$. The matter contribution that takes into account the electron density of the Earth are given by
\begin{equation}
    \hat{H}_{mat} = \pm\sqrt{2} G_f N_e \begin{pmatrix}
        1 & 0 & 0 \\ 0 & 0 & 0 \\ 0 & 0 & 0  
    \end{pmatrix}\,,
\end{equation}
where the sign depends on whether neutrinos ($+$) or antineutrinos ($-$) are involved, $G_f$ is the Fermi coupling constant and $N_e$ the electron density. In the literature, often different forms of $\hat{H}_{vac}$ are used that differ from the one presented here by the addition of a constant matrix in the mass basis and energy differences agree for all choices\footnote{See for instance \cite{Oliveira:2016asf,Carpio:2017nui,Coelho:2017byq} where this is used to work with a matrix in which one of the diagonal elements is zero:
\begin{equation}
     \hat{H}_{vac} =  \frac{c^4}{2 E}
 \begin{pmatrix}
0 & 0 & 0 \\ 0 & \Delta m_{21}^2 & 0 \\ 0 & 0 & \Delta m_{31}^2     
\end{pmatrix}
\end{equation}}. As shown in appendix \ref{secADiag}, this yields the same result as long as the final solution of the master equation only depends on energy differences $(E_i-E_j)$ or powers thereof because constant terms that agree for all $i,j$, e.g. $E$, are just canceled in the difference. While the \eqref{eq:vacmb} is formulated in terms of the vacuum mass basis, to solve the master equation it is advantageous to work in the effective mass basis in which $\hat{H}_S^{(0)}$ is diagonal\footnote{See appendix \ref{secADiag} for an efficient way of computing the diagonalisation of the matrix in question}. This basis always changes when $N_e$ changes, i.e. when we consider different layers of the Earth. We denote all quantities in the effective mass basis with a tilde and the transformation matrix with $\hat{\widetilde{V}}$. If we define the diagonal matrix of the system in the effective mass basis $\hat{\widetilde{H}}_S := \hat{\widetilde{V}}{}^\dagger \hat{H}_S^{(0)} \hat{\widetilde{V}}$, the master equation in terms of the effective mass basis can be written as
\begin{align}\label{eq:masterequeffmb}
    \frac{d}{dt}\hat{\widetilde{\rho}}_S(t) =&-\frac{i}{\hbar} \left[ \hat{\widetilde{H}}_S, \hat{\widetilde{\rho}}_S(t)\right] +\frac{8\eta^2}{\hbar^2} \frac{Tk_B}{\hbar} \left( \hat{\widetilde{H}}_S \hat{\widetilde{\rho}}_S(t) \hat{\widetilde{H}}_S -\frac{1}{2} \left\{ \hat{\widetilde{H}^2_S}, \hat{\widetilde{\rho}}_S(t) \right\} \right)\,.
\end{align}
As we have already seen from \eqref{eq:MasterLindblad}, the dissipator involves second powers of the system's Hamiltonian, as well as is linear in temperature parameter $T$ of the gravitational environment. A consequence of the second property is that there is no decoherence effect at a temperature of zero, e.g. when the environment is in a vacuum state. An implication of the first property is that such a form of the dissipator leads with respect to the effective mass basis to a decoherence term that is quadratic in the difference of the energy eigenvalues $(\tilde{H}_i-\tilde{H}_j)^2$. This can be also seen directly from the solution of the
 differential equation for $\hat{\widetilde{\rho}}_S(t)$ in \eqref{eq:masterequeffmb}, which is discussed in the appendix \ref{secA2}. In terms of the effective mass basis, this solution is given by
\begin{equation}\label{eq:solmeqeffmb}
\widetilde{\rho}_{ij}(t) =\widetilde{\rho}_{ij}(0) \cdot e^{ -\frac{i}{\hbar} \left( \widetilde{H}_i - \widetilde{H}_j \right) t - \frac{4\eta^2k_B T}{\hbar^3} \left( \widetilde{H}_i -\widetilde{H}_j \right)^2 t } \,,
\end{equation}
where $\widetilde{\rho}_{ij}(t)$ denotes the matrix elements of $\hat{\widetilde{\rho}}_S(t)$ and $\widetilde{H}_i$ denotes the elements of the diagonal matrix $\hat{\widetilde{H}}_S$. In relation to the effective mass basis, we explicitly obtain
\begin{align}
\label{eq:rhoSolRenorm}
    \rho_{ij}(t) = \rho_{mn}(0) \widetilde{V}_{km}^\dagger \widetilde{V}_{nl} \widetilde{V}_{ik} \widetilde{V}_{lj}^\dagger e^{-\frac{i}{\hbar} \left( \widetilde{H}_k - \widetilde{H}_l \right) t- \frac{4\eta^2k_B T}{\hbar^3} \left( \widetilde{H}_k -\widetilde{H}_l \right)^2 t}
\end{align}
where $\widetilde{V}_{ij}$ denote the matrix elements of $\hat{\widetilde{V}}$. The first term corresponds to the standard oscillation term which is non-vanishing for $\Delta m_{ij}^2 \not=0$. The second terms is the additional  contribution due to coupling to the gravitational environment. Note, that as discussed above due to the counter term that we introduced in \eqref{eq:CounterTerm} the final solutions does not involve a Lamb shift contribution and is thus independent of the cutoff frequency.  A model that also included a Lamb shift contribution is the one in \cite{Benatti:2001fa} where due to such a contribution the model allows neutrino oscillations to be present even if the initial mass difference vanishes. Although it is not directly obvious from the parameter in which the authors in \cite{Benatti:2001fa} encode the Lamb shift contribution (denoted as $\omega_3$ in (3.1) in \cite{Benatti:2001fa}), to our understanding the final value of this parameter depends on a choice of test function that needs to be introduced to regularise an otherwise infinite integral (see (A.14) in \cite{Benatti:2001fa}). Although the derivation in \cite{Benatti:2001fa} starts with a field theory setup, as far as we understand the derivation, carried over to the toy model presented here, such a test function would correspond to the cutoff frequency because the final value of $\omega_3$ will in general depend on the chosen test function. Thus, it looks like they obtain a shift in the neutrino energy eigenvalues that still depends on some regulator and we would expect that similar to what happens here in the toy model and in the field theoretical models in \cite{Anastopoulos:2013zya,Fahn:2022zql,Lagouvardos:2020laf,Oniga:2015lro,Blencowe:2012mp} a suitable renormalisation procedure needs to be applied to obtain a cutoff independent effect. Such an effect might be potentially non-vanishing  for the model in \cite{Benatti:2001fa} in contrast to our case since they use a different coupling to the environment but this needs to be carefully checked. To demonstrate that working with a model where no renormalisation has been applied and the Lamb shift contribution is taken as a real physical contribution, we refer to Fig.~\ref{fig:LambShift} in Sec.~\ref{sec:Results} where the toy model with and without a Lamb shift contribution are compared and thus is shown that if this non-physical effects are not removed, one might draw incorrect physical conclusions. 

\section{Results}\label{sec:Results}
Quantum decoherence in the neutrino sector has been investigated with long baseline neutrino detectors \cite{DeRomeri:2023dht}, such as MINOS+T2K \cite{minos_t2k}, the future DUNE \cite{Carpio:2018gum} and reactor experiments such as Daya Bay, RENO, and the future JUNO \cite{JUNO:2021ydg}, where the treatment of neutrino oscillations can be well approximated by the vacuum case, and neutrino telescopes such as IceCube \cite{icecube_qd, deepcore_qd} and KM3NeT \cite{Lessing:2023uxb}, where matter effects play a relevant role. All these analyses are based on a class of phenomenological quantum decoherence (PQD) models that can vary by the power with which the mean neutrino energy enters into the decoherence term, as well as by the number of non-vanishing decoherence parameters.  In order to interpret the PQD models in terms of the microscopic gravitationally induced quantum decoherence model (GQD) presented in this work, and in order to see the differences among the models, we investigate their behavior in different situations. 
~\\ ~\\
The model considered in this work has the property that only squared differences of neutrino energies enter into the terms responsible for decoherence. As a consequence this model is only compatible with a subclass of phenomenological models with an energy dependence $E^{-2}$. This follows from the specific coupling between the neutrinos and the environment inspired by general relativity and linearised gravity. Furthermore, the corresponding decoherence parameters cannot be independent from $\Delta m^2_{ij}$. Setting some of the decoherence parameters to the same value or to zero, as is done in some phenomenological models, leads to inconsistencies in the model, unless assuming the respective $\Delta m^2_{ij}$ terms are identical or vanish. 
~\\ ~\\
The results of this work show that we can obtain in the vacuum case an exact match between the aforementioned subclass of the PQD models with $n=-2$ and the GQD model presented here if we choose the decoherence parameters involved in the PQD model appropriately. It follows that for some detector configurations it is possible to set bounds to the free parameter $\eta$, which is related to the coupling constant in the interaction, and the temperature parameter $T$ of the gravitational waves, using current upper limits on the common PQD parameters $\gamma_{ij}$. However, it is important to note that all existing analyses \cite{minos_t2k,DeRomeri:2023dht,Lessing:2023uxb} make some assumptions while fitting the data, such as setting one of the $\gamma_{ij} = 0$ or two of them equal to each other, which is incompatible with the model considered here at the probability level. However, in the specific case of KamLAND \cite{DeRomeri:2023dht}, the experiment configuration allows for translating upper bounds on PQD into upper bounds on the GQD model.
~\\ ~\\
Interestingly, in the non-vacuum case such a match with the specific subclass of the PQD model cannot be achieved with those phenomenological models that assume constant decoherence parameters in each layer of the Earth, as many models do, whereas in the model presented here they do vary. As a consequence, we obtain deviations in the oscillation probabilities in matter that can become large enough in the GeV energy regime to be resolved by neutrino telescopes, such as KM3NeT/ORCA \cite{km3net_loi}. There exist models that also take matter effects in the decoherence parameters into account, see for instance \cite{Carpio:2017nui,Barenboim:2024wdn}. However, as discussed below, it is not straight forward either to match these models to the one considered in this work. Since differences in neutrino oscillations in matter represent an interesting scenario to test the model considered here in an independent way from the PQD model, all oscillation probabilities in this section are evaluated for neutrinos propagating through the Earth. Specifically, the shown two-dimensional probability distributions have been obtained with the public software package \textit{OscProb} \cite{oscprob}, where the GQD model has been implemented and added by the authors. For the Earth density profile, the \textit{PREM} model with $425$ layers has been used \cite{prem}. 

\subsection{Comparison to existing phenomenological models}
\begin{figure}[H]
\begin{minipage}{.5\linewidth}
\centering
\subfloat{\label{main:a}\includegraphics[scale=.4]{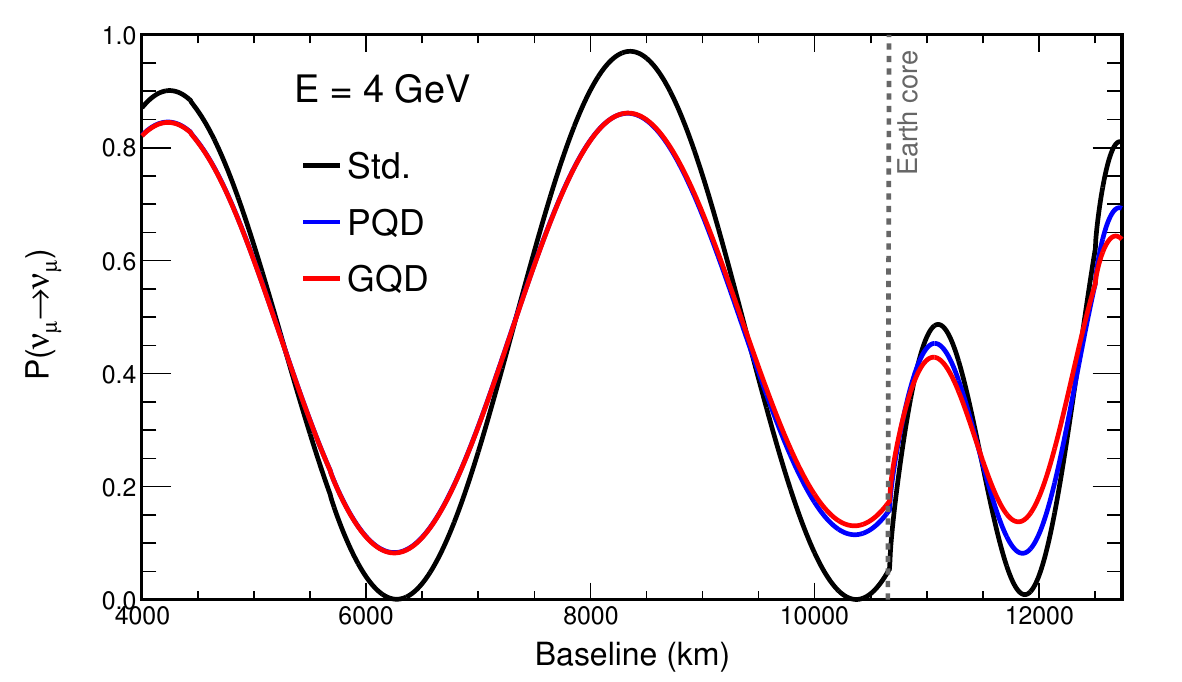}}
\end{minipage}%
\begin{minipage}{.5\linewidth}
\centering
\subfloat{\label{main:b}\includegraphics[scale=.4]{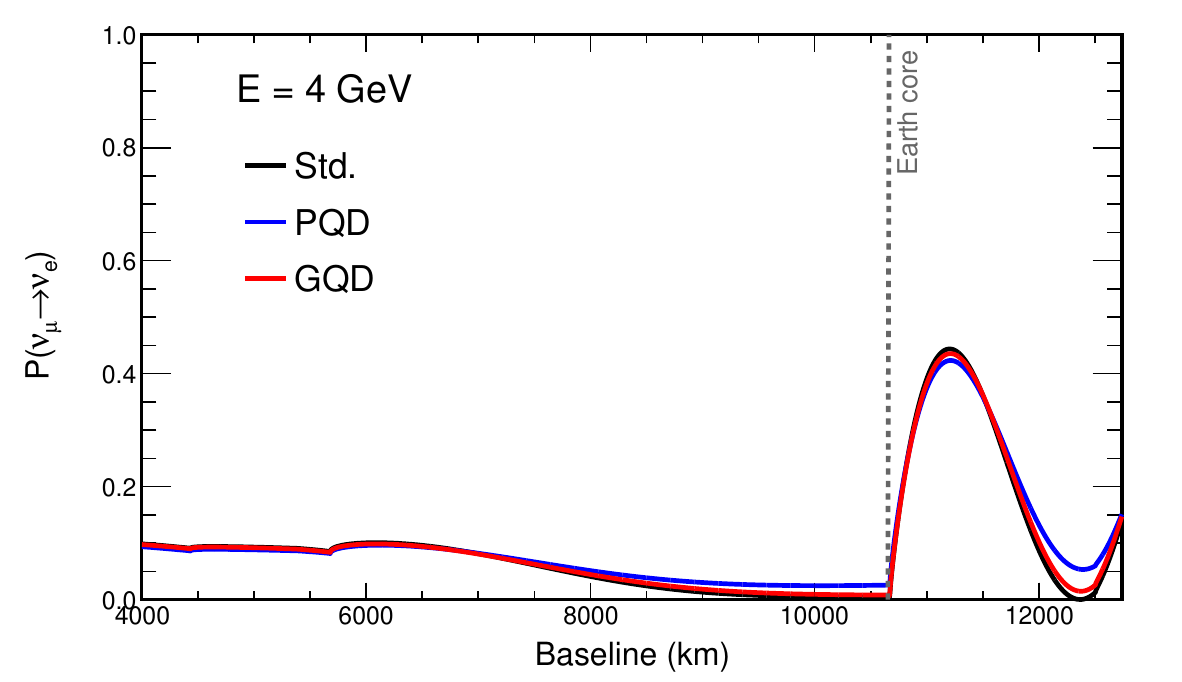}}
\end{minipage}\par\medskip
\centering
\subfloat{\label{main:d}\includegraphics[scale=.4]{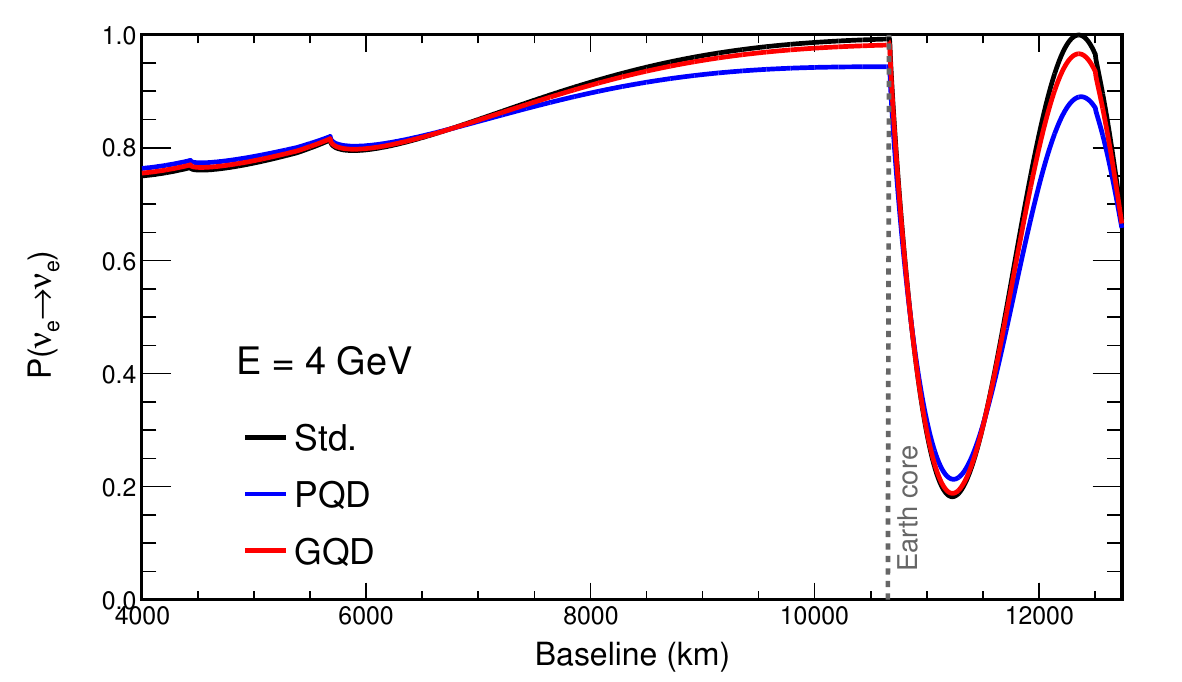}}
\caption{Neutrino oscillation probabilities in matter for a $4\,\mathrm{GeV}$ neutrino as a function of the travelled baseline in Earth. The plots compare standard oscillations without decoherence (Std.) with the model in this work (GQD) and a phenomenological model with $n=-2$ (PQD). The GQD model parameters are set to $\eta=10^{-8}\,\mathrm{s}$ and $T=0.9\,\mathrm{K}$ which implies for the PQD model in vacuum $\gamma_{21} = 1.00694\cdot 10^{-25}, \gamma_{31} = 1.08067\cdot 10^{-22}$ and $\gamma_{32} = 1.0157\cdot 10^{-22}$. The Earth matter effects are accounted for via the PREM model \cite{prem}: the clear discontinuity for neutrino trajectories passing inside of the Earth core is due to the net change in density between mantle and core. The plots have been created with \textit{OscProb} \cite{oscprob}.}
\label{fig:baseline}
\end{figure}

\begin{figure}[H]
\begin{minipage}{.5\linewidth}
\centering
\subfloat{\label{main:a1}\includegraphics[scale=.4]{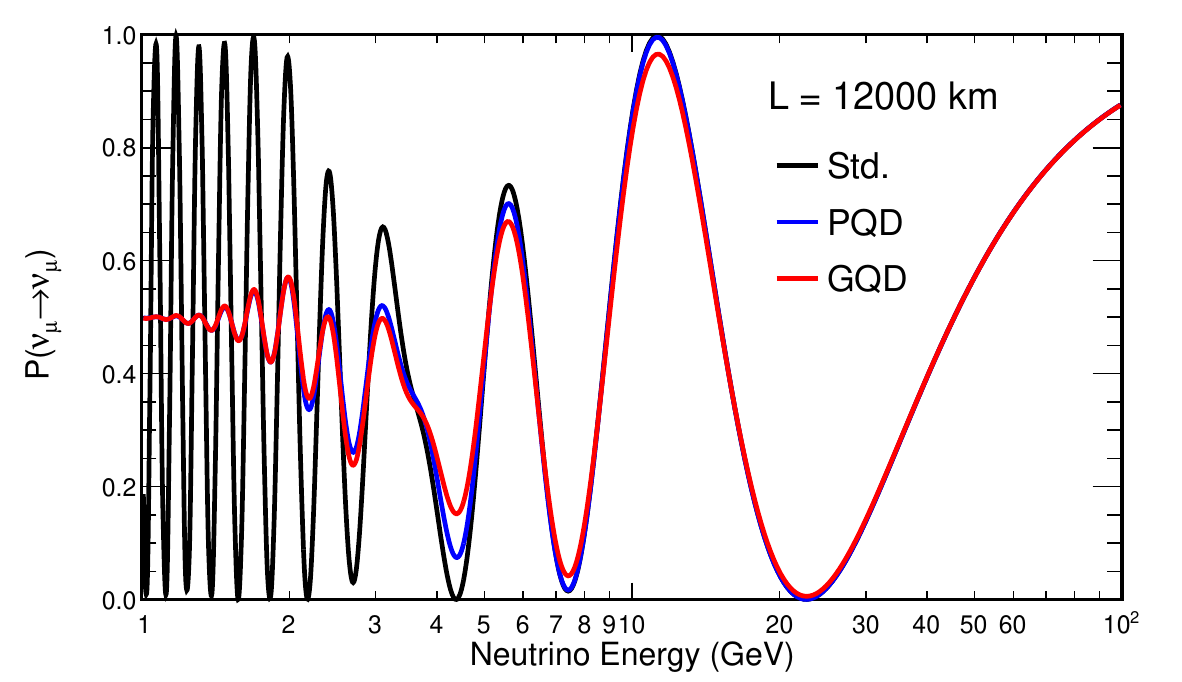}}
\end{minipage}
\begin{minipage}{.5\linewidth}
\centering
\subfloat{\label{main:b1}\includegraphics[scale=.4]{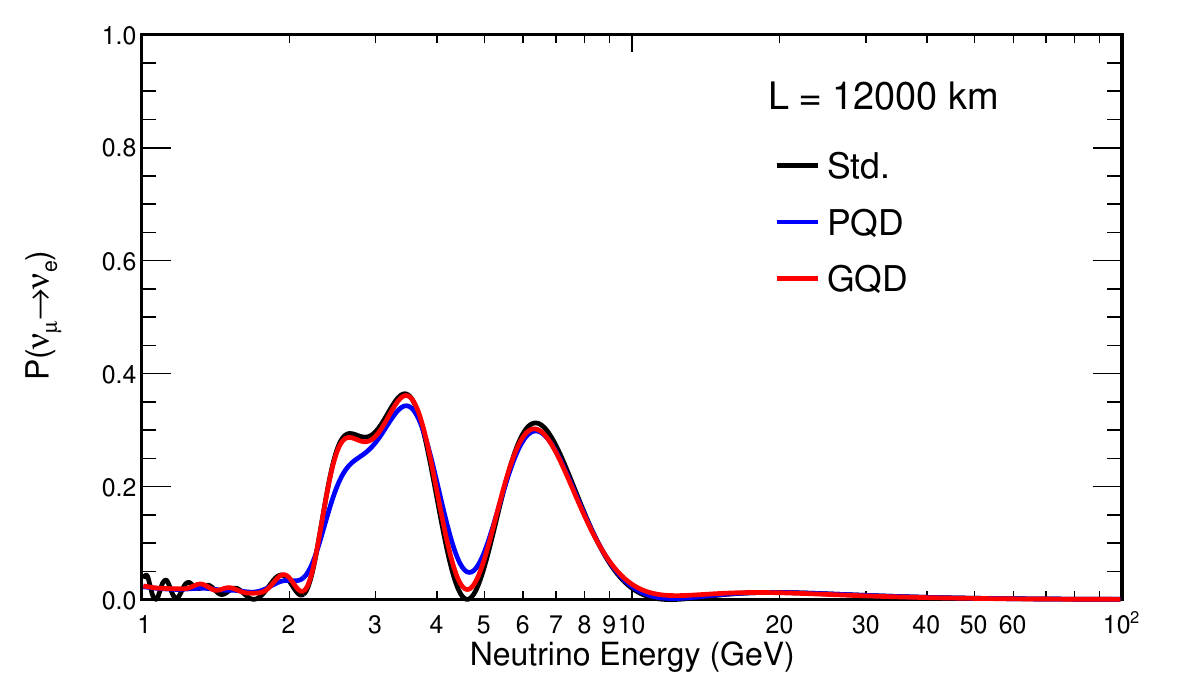}}
\end{minipage}\par\medskip
\centering
\subfloat{\label{main:c1}\includegraphics[scale=.4]{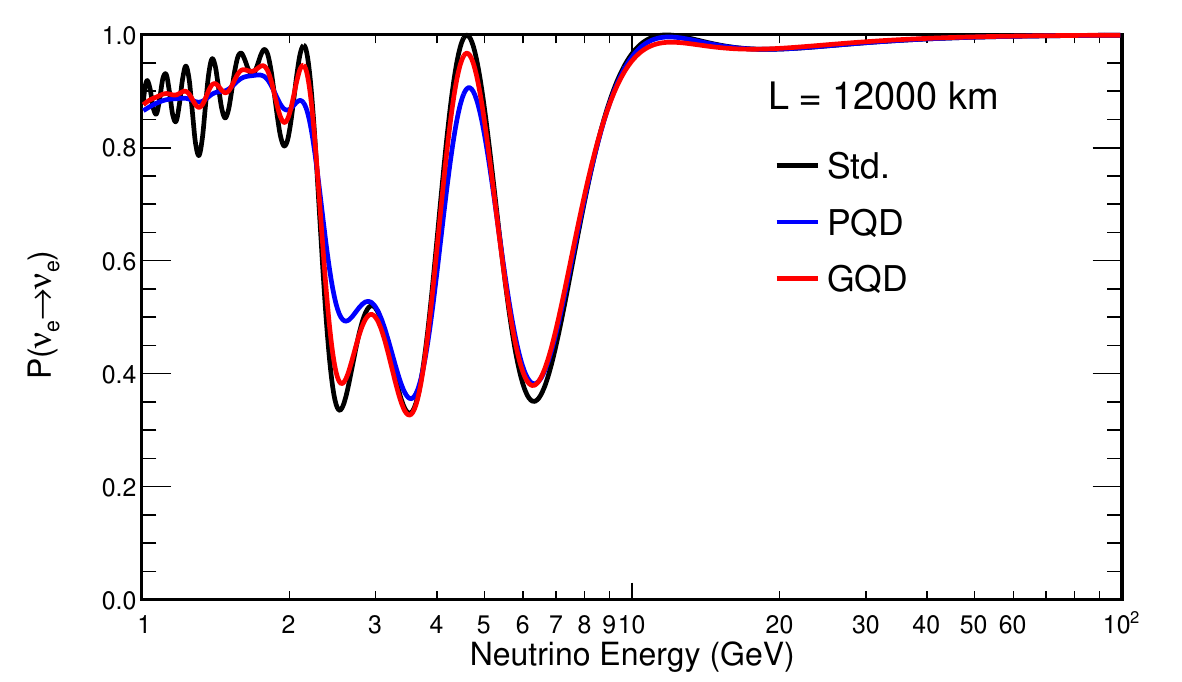}}
\caption{Neutrino oscillation probabilities in matter for a baseline of $12000\,\mathrm{km}$, corresponding to $\cos\theta_{Zenith} = -0.94$, which is a trajectory passing through the Earth core. The plots compare standard oscillations without decoherence (Std.) with the model in this work (GQD) and a phenomenological model with $n=-2$ (PQD). For the GQD model $\eta=10^{-8}\,\mathrm{s}$ and $T=0.9\,\mathrm{K}$ are assumed which implies for the PQD model in vacuum $\gamma_{21} = 1.00694\cdot 10^{-25}, \gamma_{31} = 1.08067\cdot 10^{-22}$, and $\gamma_{32} = 1.0157\cdot 10^{-22}$. The Earth matter effects are accounted for via the PREM model \cite{prem}.
The difference between constant decoherence parameters independent of the Earth matter density (PQD) and a parameter depending on the Earth matter density (GQD) is evident. The plots have been made with \textit{OscProb} \cite{oscprob}.}
\label{fig:devDecosConstGamma}
\end{figure}

\begin{figure}[H]
\begin{minipage}{.5\linewidth}
\centering
\subfloat{\label{main:a2}
\includegraphics[scale=.4]{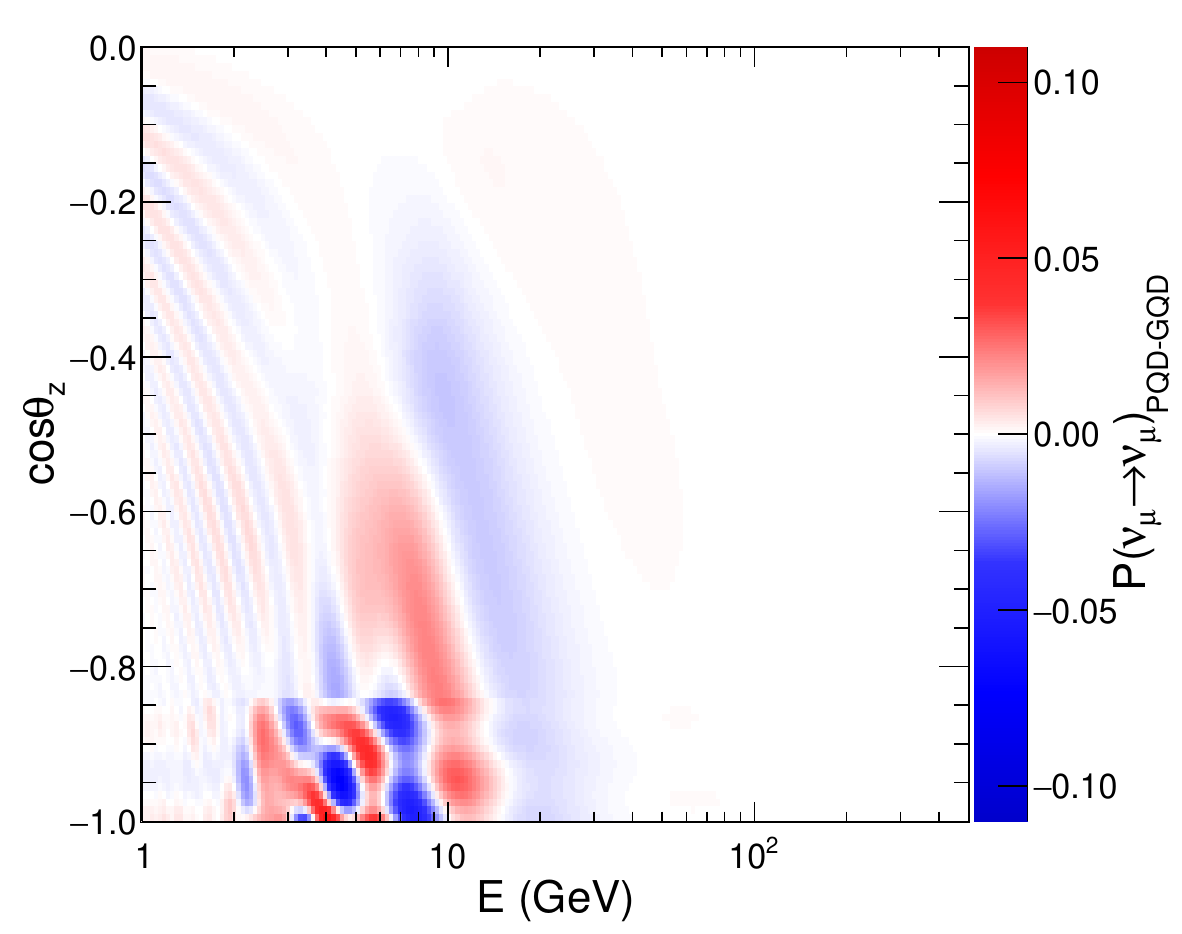}}
\end{minipage}%
\begin{minipage}{.5\linewidth}
\centering
\subfloat{\label{main:b2}\includegraphics[scale=.4]{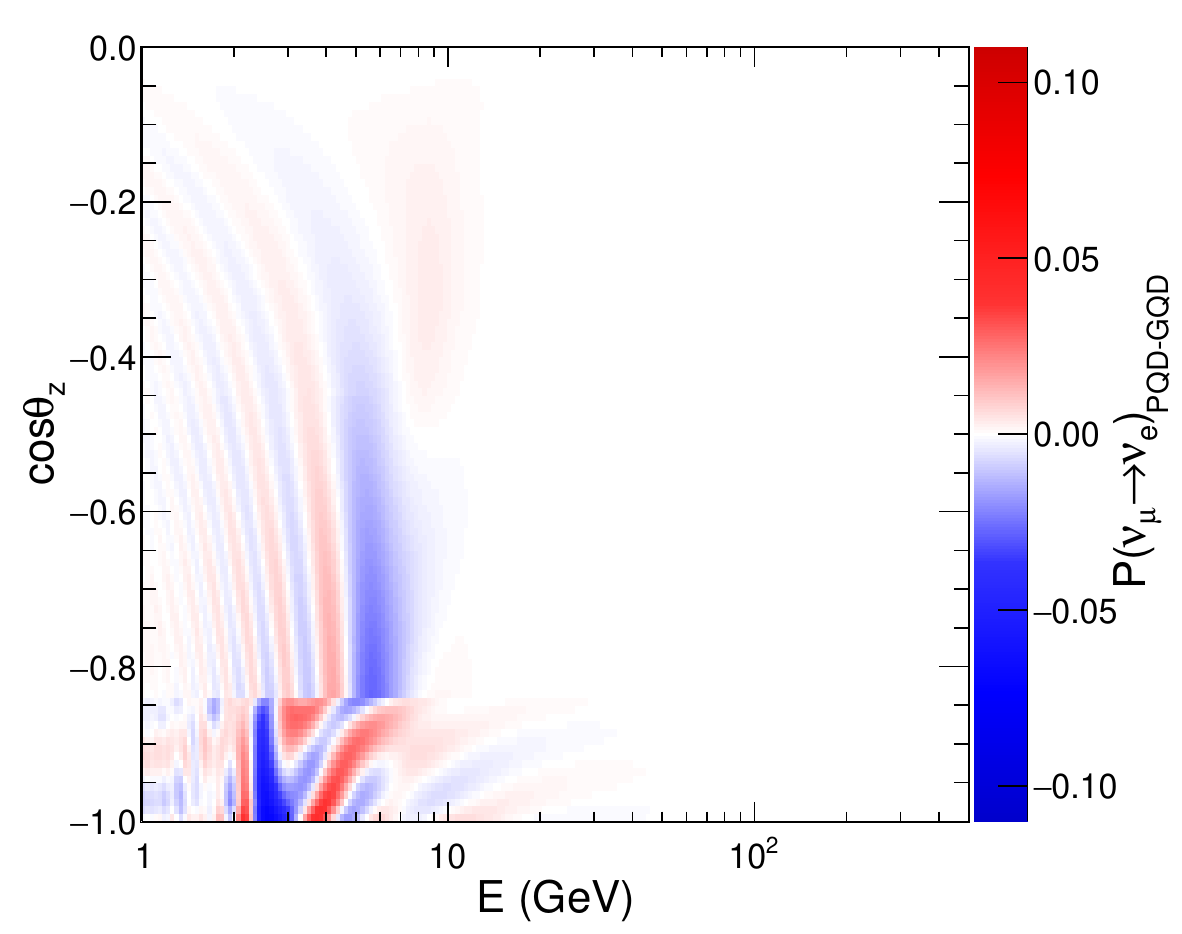}}
\end{minipage}\par\medskip
\centering
\subfloat{\label{main:c}\includegraphics[scale=.4]{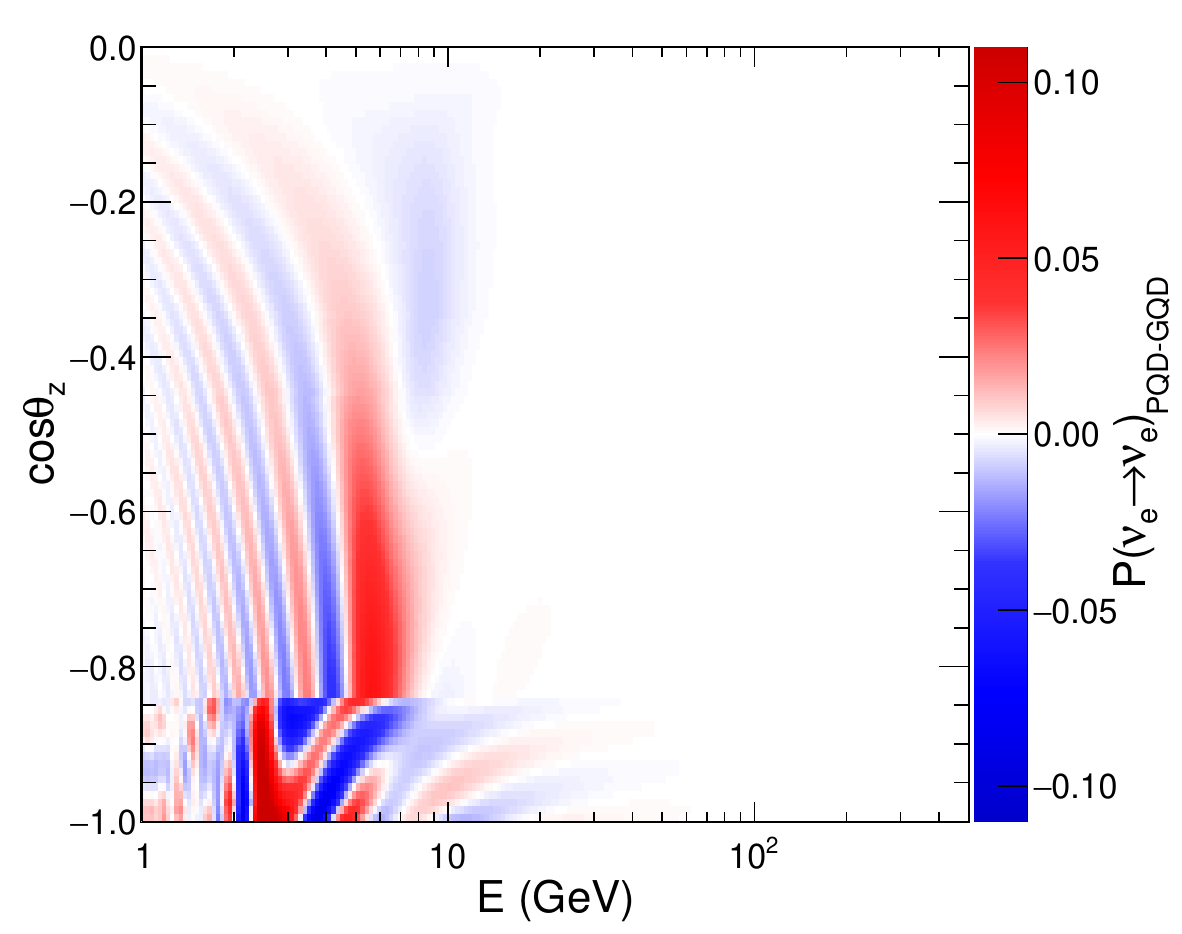}}
\caption{Difference of neutrino oscillation probabilities in matter for model in this work (GQD) and a phenomenological model for $n=-2$ (PQD). For the GQD model $\eta=10^{-8}$ s and $T=0.9\,\mathrm{K}$ are assumed which implies for the PQD model in vacuum $\gamma_{21} = 1.00694\cdot 10^{-25}, \gamma_{31} = 1.08067\cdot 10^{-22}$, and $\gamma_{32} = 1.0157\cdot 10^{-22}$. The Earth matter effects are accounted for via the PREM model \cite{prem}. The difference between constant decoherence parameters independent of the Earth matter density (PQD) and a parameter depending on the Earth matter density (GQD) is evident. The plots have been made with \textit{OscProb} \cite{oscprob}.}
\label{fig:devDecosConstGamma2}
\end{figure}

In several works, such as in \cite{Lisi:2000zt,Gago:2000qc,Guzzo:2014jbp,Coloma:2018idr,Gomes:2020muc}, a phenomenological model based on a Lindblad equation is used to model decoherence in neutrino oscillations with a solution of the form
\begin{equation}\label{eq:solmeqpheno}
\widetilde{\rho}_{ij}(t) =\widetilde{\rho}_{ij}(0) \cdot e^{ -\frac{i}{\hbar} \left( \widetilde{H}_i - \widetilde{H}_j \right) t - \Gamma_{ij} t } \,,
\end{equation}
where $\Gamma_{ij}$ is usually parameterised as
\begin{equation}\label{eq:paramGammaPhen}
    \Gamma_{ij} = \gamma_{ij} E^n\,.
\end{equation}
The model presented in this work can be related to the phenomenological model in the case of vacuum by identifying 
\begin{equation}\label{eq:connectionOurGammaPhenoGamma} 
    \gamma_{ij} = \frac{\eta^2 c^8k_BT}{\hbar^3} (\Delta m_{ij}^2)^2\quad{\rm and} \quad n=-2,
\end{equation}
where $k_B$ denotes the Boltzmann constant. Hence, the toy model considered in this work has the property that the $\gamma_{ij}$ are related to the square of the squared mass differences $\Delta m_{ij}^2$. The only two free parameters left in $\gamma_{ij}$ are $\eta$, that encodes the strength of the coupling of the neutrinos to the gravitational environment, and $T$  which is the temperature of the environment of thermal gravitational waves. If one considers cosmological models in which the usual inflationary epoch is preceded by a radiation-dominated era, a thermal gravitational wave background can be produced in the early universe. In these models, it is assumed that the thermal gravitons decouple at a temperature of the order of the Planck temperature and exhibit a black-body spectrum \cite{Kolb:1990vq,Gasperini:1993yf}. As the universe expands, the black-body spectrum of the gravitons is maintained, but the temperature is strongly red-shifted. The estimates for the temperature of the thermal gravitational wave background in the present epoch are $T\simeq 0.9\,\mathrm{K}$ \cite{Giovannini:2019oii}, and thus below the temperature of the cosmic microwave background of $T_\gamma\simeq 2.72\,\mathrm{K}$.
Furthermore, because the energy eigenvalues always involve the combination $\Delta m_{ij}^2/E$ this model suggests that the decoherence parameters depend inversely on the squared mean neutrino energy. This dependence stems from the form of the interaction Hamiltonian which was motivated by how gravity couples to matter according to general relativity. Compared to the phenomenological models, the approach presented here has the advantage that, if we assume that we obtain a value of the temperature parameter of thermal gravitational waves from other experiments, the $\Gamma_{ij}$ only depend on one free parameter $\eta$. In order to constrain the free $\gamma_{ij}$ parameters, in some phenomenological models, as for instance in \cite{Gomes:2020muc,Lessing:2023uxb,DeRomeri:2023dht}, additional requirements are included where some of the $\gamma_{ij}$ are set to zero or equal to each other. These limits then result each in  one single free parameter $\gamma$. However, from \eqref{eq:connectionOurGammaPhenoGamma} it can be seen that such choices correspond to either setting some of the $\Delta m_{ij}^2$ equal to zero or equal to each other, which stands in contradiction to experiments. Nonetheless, certain experimental configurations make these assumptions reasonable for the purpose of constraining the GQD model.
Furthermore, the physical interpretation of this parameter is harder to access compared to the situation where the underlying microscopic model is known.
\\ \\
The phenomenological models that assume constant decoherence parameters $\gamma_{ij}$ in matter cannot be matched by specific choices of parameters to the model used here. The reason for this is that while in the phenomenological models $\gamma_{ij}$ is fixed at a certain value independently of the matter density present, in the model presented in this work, the decoherence parameter depends on $( \widetilde{H}_i-\widetilde{H}_j)$, which are the energy values of the neutrino which depend on the matter density and thus the different layers in the Earth. Hence, in the microscopic model matter effects are included directly via its coupling to the environment. This dependence is caused by the fact that in matter, the vacuum Hamiltonian is extended by the additional operator $\hat{H}_{mat}$ which depends on the electron density $N_e$ in the considered matter layer, see \eqref{eq:vacmb}, and thus the final Hamiltonian and therefore also the decoherence parameter changes in each layer. Hence, the model presented here takes the effect of the different Earth layers into account in the coupling to the environment and can thus not be formulated with a single value for $\gamma_{ij}$ as it is done for the phenomenological models \eqref{eq:paramGammaPhen}. This is shown in Fig.~\ref{fig:baseline}, where a discrepancy between PQD and GQD appears when the neutrino travels through layers of increasing density within the Earth. The effect becomes relevant for neutrino trajectories passing within the Earth core. The decoherence effects considered in \cite{Carpio:2017nui} also involve contributions from the matter Hamiltonian of the neutrinos. However, it is not so simple to match these models and the one considered here as the models in \cite{Carpio:2017nui,Barenboim:2024wdn} involve only the subleading contribution of decoherence effects in order to be able to still work with analytical expressions for the oscillations probabilities and this is used to perform an analysis for DUNE and T2HK in \cite{Barenboim:2024wdn}.
~\\
In the context of the Lindblad equation the phenomenological models with constant decoherence parameters $\gamma_{ij}$ in matter can be identified with a model where the Lindblad operator is chosen to be the neutrino Hamiltonian in vacuum, whereas the model presented here chooses the full neutrino Hamiltonian as the Lindblad operator. If we restrict to the vacuum case both models obviously agree but deviate in the matter case. From the point of view of the gravitational environment it is not very obvious why the thermal gravitational background should only couple to the vacuum energy of the neutrino even if a non-vanishing matter density is present.
\\
A further consequence of this is that bounds on decoherence effects of neutrino detectors that have been derived using the PQD models can only constrain $\eta$ in the vacuum case or in the case of a single layer with constant density. Although there are upper limits on the gamma parameters from neutrino experiments where matter effects are negligible, such as the results of the MINOS+T2K data \cite{minos_t2k}, the baselines and neutrino energies of such experiments allow a vacuum treatment of neutrino oscillations. However, in \cite{minos_t2k}, they fit the data assuming three scenarios which are not fully compatible with our choice of parameters. At the same time, the KamLAND \cite{DeRomeri:2023dht} detector configuration allows for a treatment with a single layer of constant density (the Earth crust). Additionally, the energy range of its neutrino beam makes the standard assumptions used in PQD models reliable for constraining the parameters of the GQD model.
\\
For the matter case with multiple layers of different densities a further sensitivity analysis is needed because the deviations in the oscillation probabilities predicted when using the PQD model with a decoherence parameter independent of the matter density or the one presented here (GQD) is a measurable effect as can be seen in Fig.~\ref{fig:devDecosConstGamma}, where the constant $\gamma_{ij}$ were chosen to be equal to the decoherence parameters in the GQD model in vacuum such that the (PQD) and the (GQD) model perfectly match in the vacuum case. From the corresponding oscillograms in Fig.~\ref{fig:devDecosConstGamma2} it becomes visible that the probabilities can deviate by up to $10\%$ for $\eta= 10^{-8}\,\mathrm{s}$ and $T=0.9 \,\mathrm{K}$.

~\\
Figure \ref{fig:devDecosConstGamma2} shows the difference in neutrino oscillation probabilities for the PGD and GQD models in Earth as a function of the neutrino energy and cosine zenith. Up-going events have $\cos\theta_Z = -1$. The choice of the $\eta$ value matches the PQD $\gamma_{ij}$ values in vacuum near to bounds. As it can be seen, differences up to $10\%$ can be observed, highlighting the different energy behaviour of the two models which arise with matter effects. It follows that the model considered here can be independently constrained with respect to the PQD model by neutrino telescopes optimised for the GeV energy range, such as KM3NeT/ORCA \cite{km3net_loi}.

\subsection{Assessing Upper Limits from Present Constraints on Phenomenological Models}
\begin{figure}[H]
\centering
\includegraphics[width=0.6\textwidth]{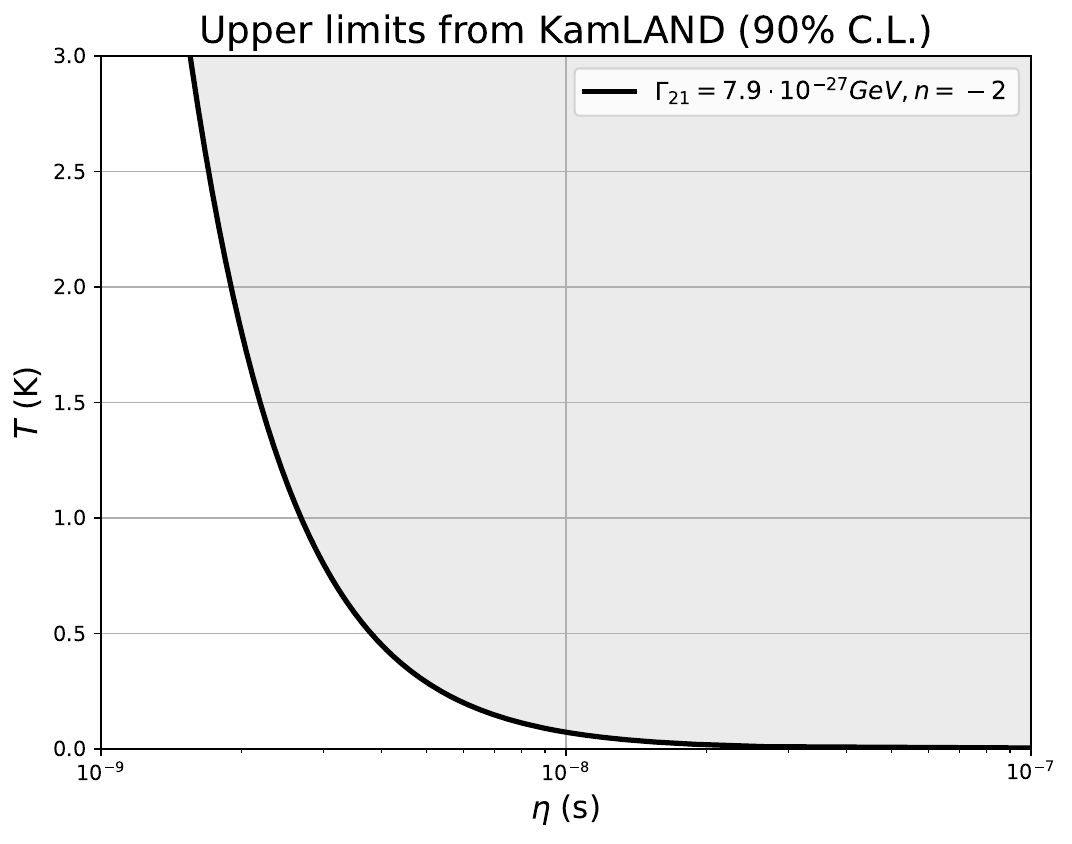}
\caption{Upper limits at $90\%$ C.L. for neutrino coupling $\eta$\,(s) as a function of the temperature $T$\,(K) of the thermal gravitational wave background. These limits have been obtained through the upper bound on \textit{Model E} investigated in \cite{DeRomeri:2023dht}. The grey region represents the excluded area.}
\label{fig:KamLAND_UL}
\end{figure}

Ref.~\cite{DeRomeri:2023dht} conducts an extensive search for quantum decoherence effects in neutrino oscillation data, employing a phenomenological approach. Among the models they examine, \textit{Model E} is of particular interest to this work. In this \textit{Model E}, the authors set $\Gamma_{31} = \Gamma_{32} = 0$ and fit $\Gamma_{21}$. 
From a pure oscillation perspective, this model would be incompatible with the model in this work, as it implies $\Delta m_{31}^2 = \Delta m_{32}^2 = 0$ (see \eqref{eq:connectionOurGammaPhenoGamma}). However, when applied to specific experiments, the model can still be employed to extract constraints in the context of this work. This is particularly true for the setup of KamLAND experiment, for which the inclusion of matter effects is necessary due to its specific baselines (up to $\sim 1000\,\mathrm{km}$) and neutrino energy range ($1.8–8\,\mathrm{MeV}$). In the KamLAND $L/E$ range, the fast oscillations driven by $\Delta m_{31}^2$ and $\Delta m_{32}^2$ get averaged out. As a result, the corresponding decoherence terms are effectively removed from the observable oscillation probability, as indicated by \eqref{eq:connectionOurGammaPhenoGamma}. The remaining decoherence effect is then associated with the $\Delta m_{21}^2$ oscillation term. Consequently, the KamLAND bound reported in \cite{DeRomeri:2023dht} can be translated into constraints on $\eta$ and $T$ using \eqref{eq:connectionOurGammaPhenoGamma}. For the $n = -2$ case, it is noteworthy that the KamLAND bounds are the most stringent among those analyzed in \cite{DeRomeri:2023dht}. This thus allows as well to impose the most stringent constraints on the parameters of the model in this work. Specifically, with a $90\%$ C.L. upper limit on $\Gamma_{21}$ of $7.9\cdot10^{-27}\,\mathrm{GeV}$ for $n=-2$, we derive the upper bounds on $\eta$ and $T$ as presented in Fig.~\ref{fig:KamLAND_UL}. 

\subsection{Effect of the renormalisation}
\label{sec:ResRenorm}
\begin{figure}[H]
\centering
\includegraphics[width=0.6\textwidth]{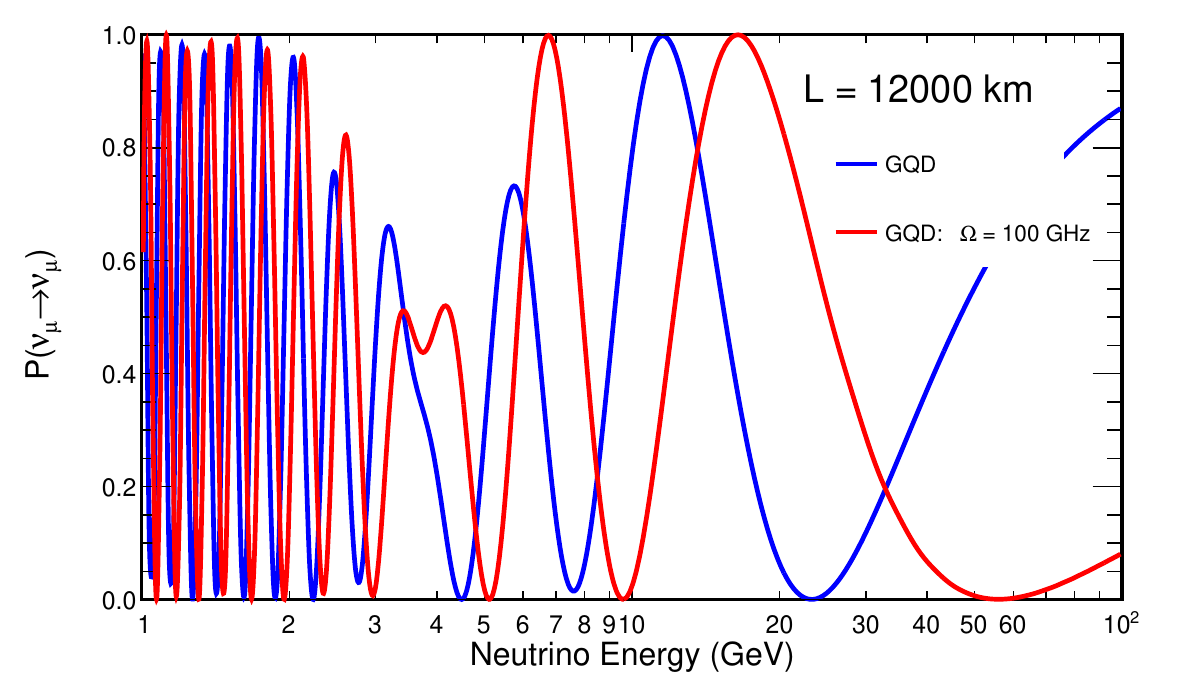}
\caption{Comparison of the model presented in this work (GQD) with and without renormalisation. In the latter case the cutoff-dependent Lamb shift contribution is still present in the unrenormalised Hamiltonian. The cutoff frequency has been chosen to be $\Omega=100\,\mathrm{GHz}$ corresponding to the maximal frequency of thermal gravitational waves at temperature $T=0.9\,\mathrm{K}$, see for instance \cite{Giovannini:2019oii} and $\eta=10^{-37}$ s for which the oscillation with $\Omega=100\,\mathrm{GHz}$ do not get too fast. The value $\Omega=100\,\mathrm{GHz}$ was chosen because it corresponds approximately to the maximal frequency of the black-body spectrum for $T=0.9\,\mathrm{K}$. Considering the Lamb shift as a real physical effect is problematic because its contribution still depends on the cutoff frequency $\Omega$ and diverges when $\Omega$ is sent to infinity, which shows that a renormalisation procedure is required. The plot has been made with \textit{OscProb} \cite{oscprob}.
\label{fig:LambShift}}
\end{figure}

In Fig.~\ref{fig:LambShift} a comparison of the model presented in this work (GQD) with and without renormalisation is presented.
As expected in the GQD model the contribution of the Lamb shift in the non-renormalised Hamiltonian $H_S^{(0)}$ in \eqref{eq:MicroHam} leads to an energy-dependent phase shift in the oscillations. However, as discussed in Sec.~\ref{sec:MicrosModelNeutrino}, such a phase shift is non-physical because it still depends on the chosen cutoff frequency $\Omega$ and diverges in the limit $\Omega\to\infty$. After renormalising the neutrino Hamiltonian and considering the limit value $\Omega\to\infty$, the contribution of the Lamb shift is not present, as it is exactly cancelled by the counter term introduced in \eqref{eq:CounterTerm}. Thus, these results show on the one hand that for the model considered in this work it is a justified procedure to do not consider the lamb-shift as well as the counter term at the level of the Lindblad equation, which is often done but needs in general a detailed analysis for each individual model separately. On the other hand, as already discussed in the context of massless neutrino oscillations discussed in \cite{Benatti:2001fa}, it further demonstrates that any physical interpretation of effects caused by the lamb shift term without a detailed analysis of the renormalised model can be problematic

\subsection{Coupling strength inspired from linearised gravity}
\begin{figure}[H]
\centering
\includegraphics[width=0.6\textwidth]{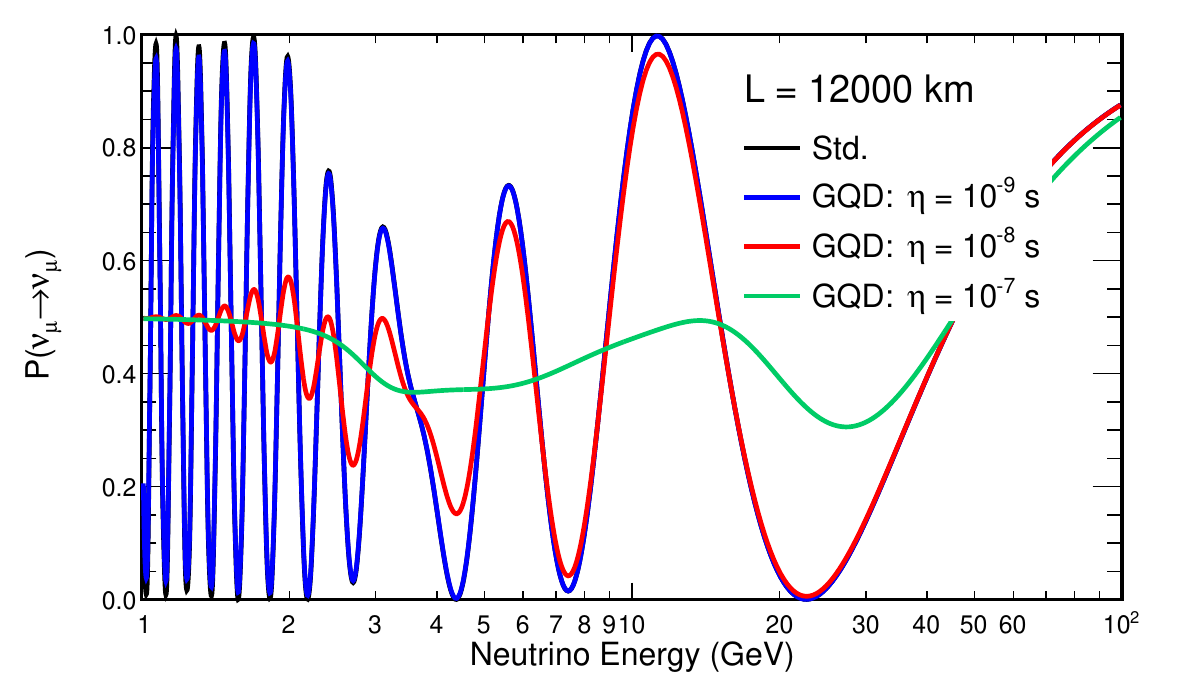}
\caption{Effect of the neutrino coupling to the gravitational environment, $\eta$, on the muon neutrino disappearance probability in Earth. Here, $T = 0.9\,\mathrm{K}$ is assumed. The considered baseline assumes a neutrino propagation that passes through the Earth core. The plot has been made with \textit{OscProb} \cite{oscprob}.}
\label{fig:etas}
\end{figure}

In the model studied in this work, $\eta$ is a free parameter which represents the coupling strength between the neutrinos and the gravitational environment. It cannot further be specified by the microscopic model in \eqref{eq:MicroHam}, as it depends on the $g_i$ which are in turn not further specified. As discussed, the existing constraints from experimental data in \cite{minos_t2k,Lessing:2023uxb,DeRomeri:2023dht} cannot be used to further constrain $\eta$ because the assumptions used for the decoherence parameter in this analysis are not compatible with the model considered in this work. Hence, a further investigation on neutrino detectors sensitivities is needed to obtain such upper bounds. 
~\\ ~\\
From the theoretical point of view, the estimated value of $\eta$ is determined by the way general relativity couples to matter. However, to the knowledge of the authors no full field-theoretic model for neutrinos has yet been derived, so there is yet no definite answer to the size of $\eta$. By comparison to full field-theoretic models like \cite{Fahn:2022zql,Anastopoulos:2013zya,Lagouvardos:2020laf} it is nevertheless possible to attempt a first rather naive estimate for a suitable order of magnitude for the coupling parameter $\eta$. Using the model from \cite{Fahn:2022zql}, such an estimate can be found in appendix \ref{secA1}. The resulting value $\eta \approx 10^{-42}\,\mathrm{s}$ is rather tiny and corresponds to a $\gamma_{ij}$ in the phenomenological models (for $n=-2$) of the order $\gamma_{ij} \approx 10^{-50} \frac{\mathrm{eV^2}}{\mathrm{s}}$. For such a tiny value of $\eta$ modifications in the probability for the neutrino oscillations for similar values of the other involved parameters that have been used in Sec.~\ref{sec:Results}  one would not be able to detect modifications from the standard neutrino oscillations. This can also be seen from Fig.~\ref{fig:etas} where the modification in the probability for the neutrino oscillations are shown for three different values of $\eta$.
~\\
Specifically, for $\eta=10^{-9}$ s, the GQD model is already almost not distinguishable from the standard scenario. For $\eta=10^{-8}$ s the modifications start to become non-negligible and they become large already at $\eta=10^{-7}$ s, which corresponds, in vacuum, to values for the $\gamma_{ij}$ parameters of the PQD models of the order $\gamma_{ij} \approx 10^{-21} \frac{\mathrm{eV^2}}{\,\mathrm{s}}$. 
\\ 
Does this mean that, taking this estimate seriously, the effect of gravitationally induced decoherence will be too tiny? The answer to this question is not so simple and under debate in the current literature. For instance in \cite{Anastopoulos:2013zya,Lagouvardos:2020laf} it is discussed that the interpretation of the temperature parameter $T$ as the temperature of the thermal gravitational waves is too restrictive. They also conclude that for $T=0.9\,\mathrm{K}$ the $\eta$ that follows from the QFT model is too small to cause detectable decoherence effect. However, they argue that $T$ should rather be interpreted as an effective parameter that for $T=0$ includes gravitons in a vacuum state where no decoherence effects are present. For the choice of $T\simeq 0.9\,\mathrm{K}$\footnote{In \cite{Anastopoulos:2013zya,Lagouvardos:2020laf} they choose $T=2.7$K, the temperature of the cosmic microwave background. To the authors' knowledge, the temperature of thermal gravitational waves is expected to be somewhat lower, see e.g. \cite{Giovannini:2019oii}.}, this corresponds to cosmic thermal gravitational waves. For higher values of an effective $T$ parameter they suggest that it can for instance be given if one chooses a quantum state for the environment that mimics a classical stochastic noise with an astrophysical origin such as a background caused by all rotating neutron stars in the galaxy. Another possibility they discuss is that, assuming that classical spacetimes arise from an underlying theory of quantum gravity at the thermodynamic level, a classical spacetime such as flat Minkowski spacetime is a macrostate and its emergence is therefore accompanied by classicalised thermodynamic fluctuations, which may be stronger than any quantum fluctuations in perturbative quantum gravity. 
\begin{figure}[H]
\centering
\includegraphics[width=0.6\textwidth]{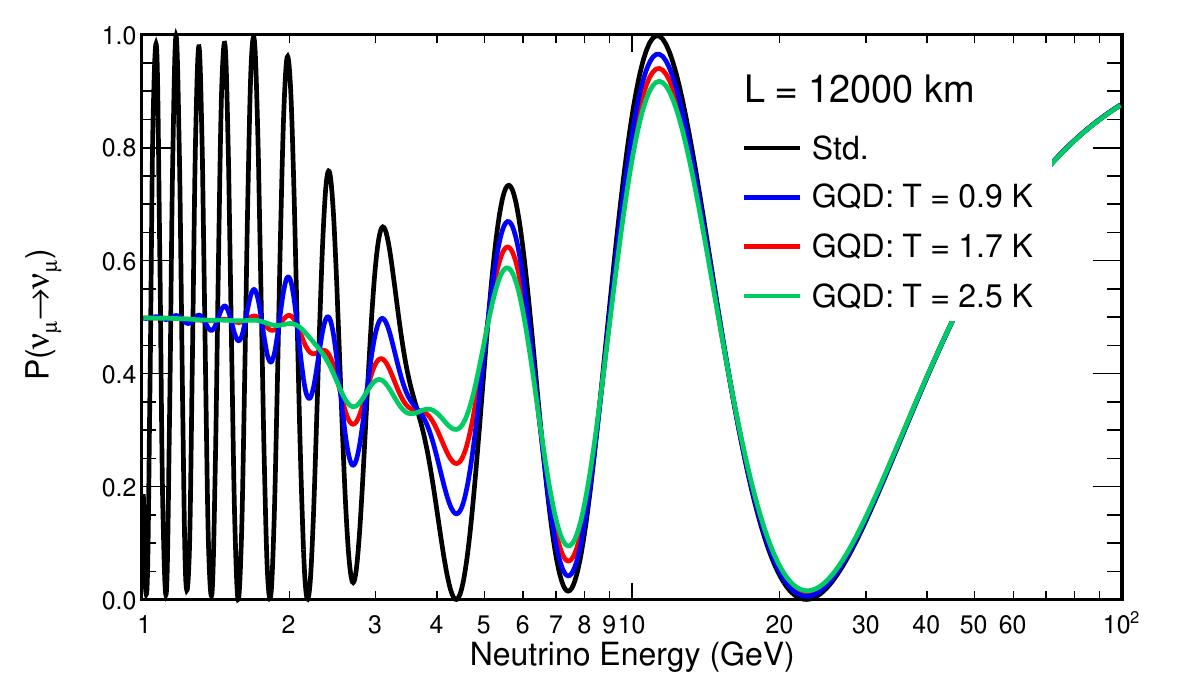}
\caption{Effect of the temperature of the gravitational environment $T$ on the muon neutrino disappearance probability in Earth. Here, $\eta = 10^{-8}\,\mathrm{s}$ is assumed. The plot has been made with \textit{OscProb} \cite{oscprob}.}
\label{fig:temperature}
\end{figure}
~\\
Therefore, $T$, which is understood as an effective parameter in their discussion, cannot be determined by a QFT model based on linearised gravity. Further decoherence models with a similar parameter involved can be found in \cite{Milburn:1991zkb,Milburn:2003zj,Breuer:2008rh,Diosi:2004iq}, where either the parameter is not further specified or the value of the Planck temperature is discussed which is an obvious but not very restrictive upper bound for such an effective temperature parameter. To address this point Fig.~\ref{fig:temperature} shows the effect of the temperature of the gravitational environment for neutrino oscillation probabilities in Earth as a function of the neutrino energy, for a fixed value of $\eta$. As it can be seen, a small variation in the temperature has visible effects in oscillation probabilities which could potentially be resolved by neutrino telescopes such as KM3NeT/ORCA. 
\begin{figure}[H]
\begin{minipage}{.5\linewidth}
\centering
\subfloat{\label{main:A2}
\includegraphics[scale=.4]{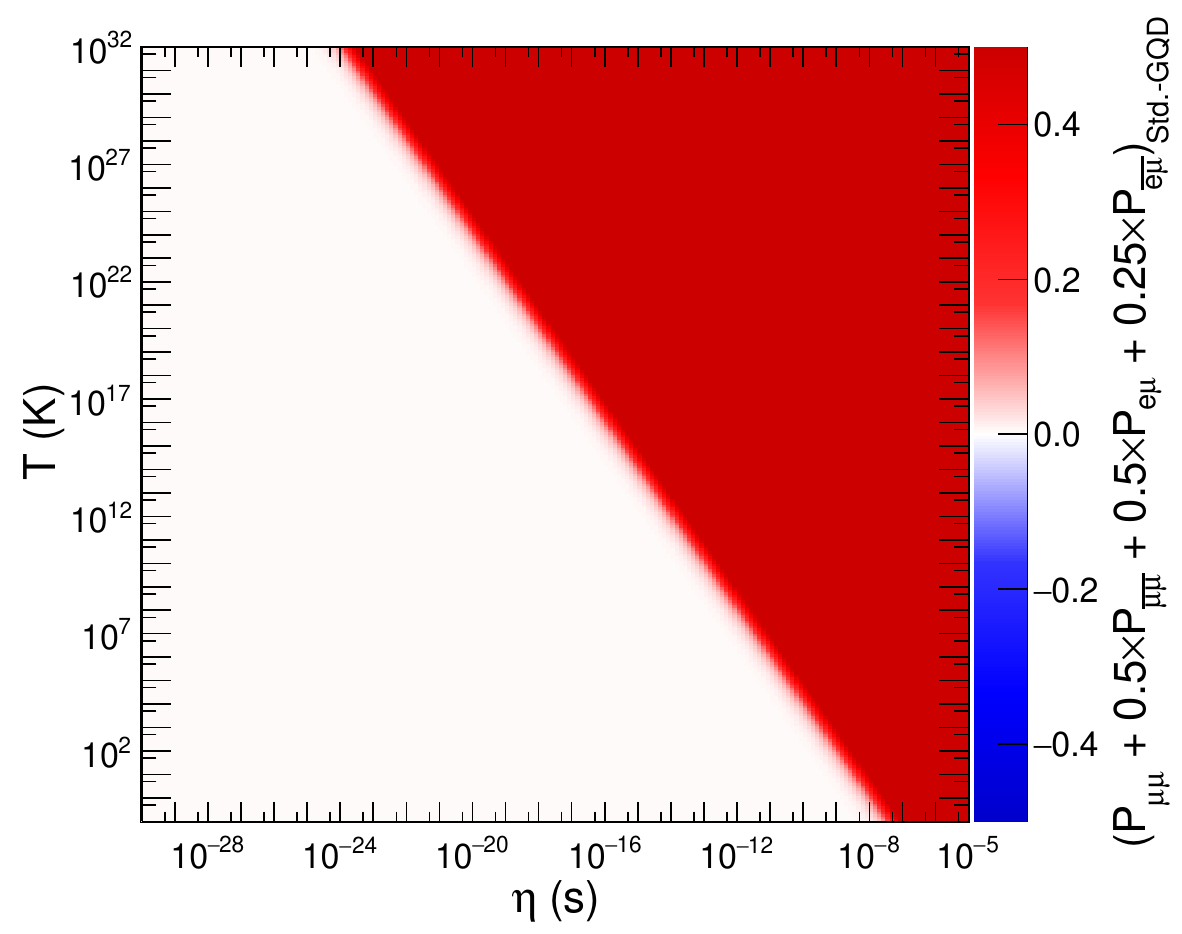}}
\end{minipage}%
\begin{minipage}{.5\linewidth}
\centering
\subfloat{\label{main:B2}\includegraphics[scale=.4]{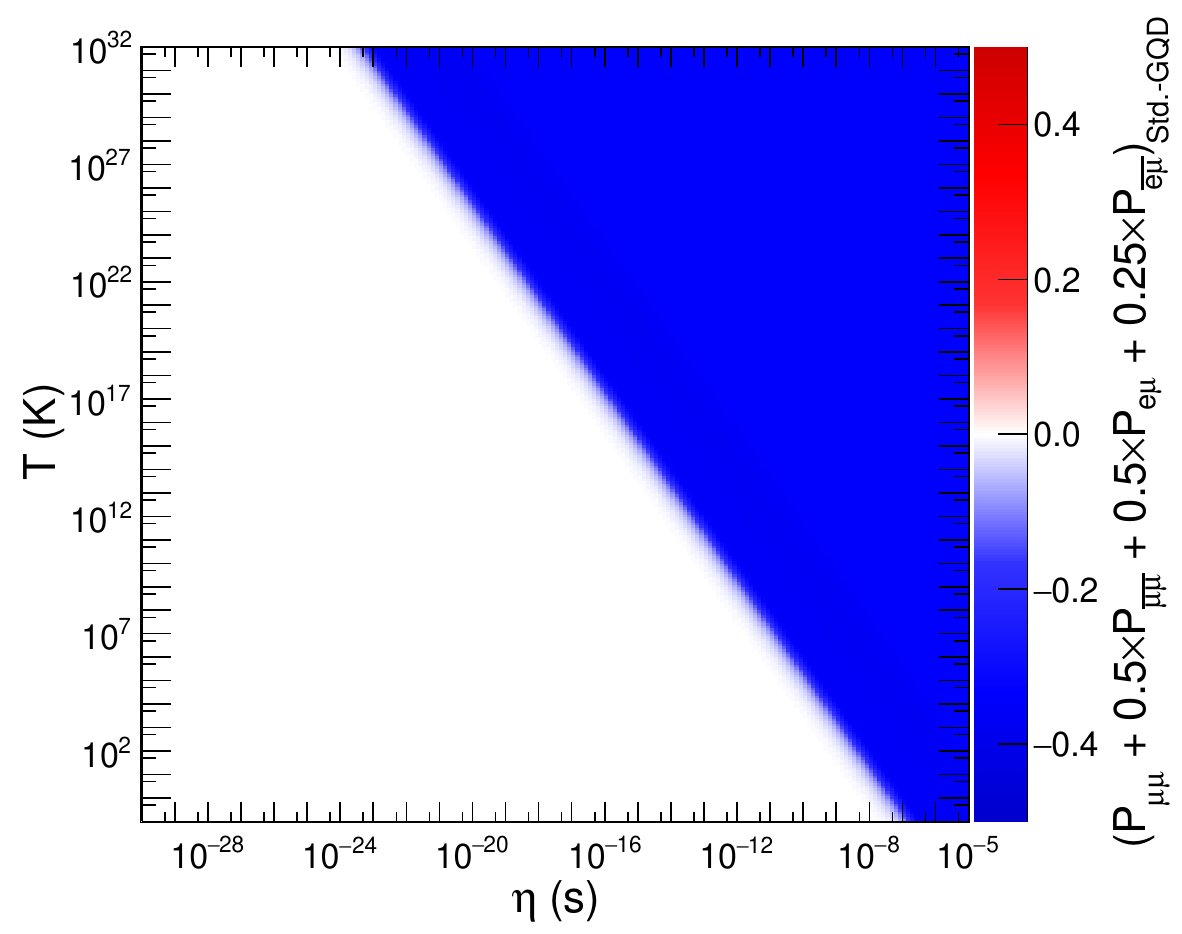}}
\end{minipage}\par\medskip
\caption{Difference in atmospheric neutrino oscillation probabilities between the standard (Std.) scenario and GQD, as a function of $\eta$ and $T$, the latter starting at $0.1\,\mathrm{K}$ and going up to the Planck temperature. Here, upgoing neutrinos ($\cos\theta_{\mathrm{Zenith}} = -1.0$) and with an energy of $6\,\mathrm{GeV}$ (left) and $18\,\mathrm{GeV}$ (right) are considered. The assumed ratio between neutrino flavours and neutrinos vs antineutrinos is a simplified approximation of the atmospheric neutrino flux. Matter effects are included via the PREM model \cite{prem}. The plots have been made with \textit{OscProb} \cite{oscprob}.}
\label{fig:eta_vs_t}
\end{figure}
~\\
In addition we show in Fig.~\ref{fig:eta_vs_t} exemplary how the modifications of the probabilities for the oscillations vary with the temperature for atmospheric neutrinos. As expected and can be seen in Fig.~\ref{fig:eta_vs_t} for higher temperatures also the parameter $\eta$ can be smaller and still deviations in the oscillations probabilities are present. The comparison between the two plots in Fig.~\ref{fig:eta_vs_t} show that for larger smaller energies the values of $\eta$ from which on deviations to the standard oscillations are seen can be slightly smaller which is due to the fact that the mean neutrino energy enters with an inverse second power in the decoherence term. As for this plots atmospheric neutrinos with the maximum propagation length through the Earth were considered the analysis of the $T,\eta$ range is in this sense restricted as we expect decoherence effects to become larger for larger propagation length.
~\\ ~\\
A further aspect related to this that becomes relevant in the case of neutrino oscillations is that we can also enhance the decoherence effect if we consider longer propagation length of the neutrinos. In this case, where one wants to consider cosmic neutrinos, for example, a QFT model based on linearised gravity as in \cite{Blencowe:2012mp,Anastopoulos:2013zya,Lagouvardos:2020laf,Fahn:2022zql} around a flat Minkowski spacetime might be too simple, and one would have to consider more complex models that involve gravitational waves in a FLRW spacetime as a more realistic setting, for which the estimate is also expected to change due to the presence of a non-trivial scale factor in cosmological spacetimes. From the point of view of quantum gravity, it therefore remains an exciting question whether decoherence effects in neutrino oscillations are actually measured and, if so, which theoretical models can then satisfactorily explain such measurements.

\section{Conclusion}\label{sec:Concl}
In this work, we have investigated gravitationally induced decoherence for neutrino oscillations based on the specific microscopic toy model in \cite{Xu:2020lhc}, which we have slightly generalised to apply it in the context of neutrino oscillations. As any open quantum system, the model includes the choice of the system, here neutrinos, and its environment, which is modelled by a finite number of harmonic oscillators that mimic the thermal gravitational waves in the toy model in  \cite{Xu:2020lhc}. In addition, we specify the coupling between system and environment as strongly inspired by the way how gravity couples to matter in the toy model \cite{Xu:2020lhc} and thus guided by the existing field theory models \cite{Anastopoulos:2013zya,Fahn:2022zql,Lagouvardos:2020laf,Oniga:2015lro,Blencowe:2012mp}.
~\\ ~\\
Our results give new insights on the physical properties of the underlying theoretical model and this allows us also to get a deeper understanding of the relation to the existing phenomenological models as well as the physical implications of the latter.
Our analysis also allows us to take up some points that are debated in the literature and look at them from a different angle.
~\\ ~\\
This first concerns the work in \cite{DEsposito:2023psn}, where the model from \cite{Xu:2020lhc} is also considered in a two-neutrino scenario, but the conclusion is drawn that no decoherence effects occur for this model when they apply the equal-energy condition. In contrast, the results in the earlier work in \cite{Adler:2000vfa}, where a similar contribution in the decoherence terms was obtained for the special case of vacuum oscillations, as well as our results show decoherence effects where both models apply the equal-momentum condition in the quantum mechanical context using plane waves. The four possibilities of equal-momentum, equal-energy, equal-velocity or conservation of energy-momentum in the plane wave approach, for which it seems rather problematic to establish these conditions on solid ground, lead to the same probabilities of neutrino oscillations, at least in the relativistic limit, if no decoherence effects occur \cite{Kruppke:2007iwa}. Furthermore, as shown in \cite{Giunti:2004yg} if one derives the formula for the neutrino oscillation probabilities from QFT, it agrees with the result obtained by using the equal-momentum condition.
~\\
 The use of the equal-energy condition in the paper in \cite{DEsposito:2023psn} is to our understanding motivated by the wave packet approach, where the derivation of the probabilities for neutrino oscillations includes an average over the detection time. For standard neutrino oscillations without decoherence effects caused by some coupling to the environment, such an average over the detection time leads to a delta function that is compatible with the application of the equal-energy condition \cite{Akhmedov:2009rb}. 
~\\ 
Since the derivation of our master equation in Sec.~\ref{sec:MicrosModelNeutrino} is done using plane waves, the model considered in this work does not include decoherence effects caused by considering wave packets instead of plane waves, and we apply the equal-momentum condition in the model.  A complete generalisation of the model presented in this paper within the wave packet approach is beyond the scope of this paper. However, from the derivation with plane waves, we can already see that the average over the detection time becomes more subtle when decoherence effects due to coupling with an environment are present. Firstly, the expression for the probability of neutrino oscillations in the presence of decoherence effects includes an additional exponential time-dependent damping term. Therefore, the integrand relevant to the average over the detection time changes and thus the result is no longer simply a delta function. Since the average over the detection time is usually performed as an integral from $-\infty$ to $+\infty$, the exponential decay for negative values of the time coordinate appears problematic, as also discussed in \cite{DEsposito:2023psn}.  As discussed in \cite{Guff:2023xzf}, an interesting question is whether there is a symmetry of time reversibility in open quantum systems. For the Lindblad equation, they show in \cite{Guff:2023xzf} that it can be extended to the negative real time axis if the absolute value for the time coordinate in the dissipator term is taken into account. This has two effects: First, in this case the average over the detection time can be formed, and second, the relevant integral for the kind of models considered here and in \cite{DEsposito:2023psn} can be solved analytically, leading to a Lorentz function instead of a delta function. Thus, if decoherence effects are present in the wave packet approach, the argument with the delta function motivating the use of the equal-energy condition cannot be transferred exactly from the situation in which no decoherence effects are present to the situation in which decoherence effects are involved. Furthermore, the exact shape of the Lorentz function depends on the particular decoherence model, and thus the final result of the average over the detection time as well. We plan for future work to investigate more in detail how the limit of the plane wave approach can be rediscovered from the wave packet approach in the case where such environmentally induced decoherence effects are present to be able to compare these different models more in detail.
~\\ ~\\
Secondly, as discussed in Sec.~\ref{sec:MicrosModelNeutrino}, it was necessary to perform a renormalisation of the neutrino Hamiltonian as otherwise we end up with a final decoherence model that still depends on a cutoff frequency of the thermal gravitational wave environment and, which is even more problematic, the non-renormalised model involves divergences if the cutoff goes to infinity. The introduction of the cutoff frequency was necessary to regularise an otherwise infinite integral over the spectral density in the derivation of the master equation. In the non-renormalised Hamiltonian, the cutoff frequency is included in the so-called Lamb shift contribution, which shifts the energy eigenvalues of the neutrinos in a way that depends on the cutoff frequency and the energy. After the renormalisation, this shift of the energy eigenvalues is no longer present, so that the Lamb shift makes no contribution and, in our understanding, can only be interpreted as an unphysical effect before renormalisation. This seems to be in contrast to the interpretation used in \cite{Benatti:2001fa}, where an analogous contribution of the Lamb shift, which in their case depends on a test function as a regulator, is used to explain massless neutrino oscillations. From our point of view, one would also have to apply a renormalisation procedure which we expect to lead to no contribution from the Lamb shift. However, there might be additional finite terms that survive after renormalisation, as their occurrence crucially depends on the coupling between system and environment, which is chosen differently in \cite{Benatti:2000ph} than in our case, but such a conclusion requires further investigation. 
~\\ ~\\
In addition, even if we consider the non-renormalised version of the model presented in this work, no massless neutrino oscillations will be allowed by the model. The reason for this is that in the model considered here the Lamb shift contribution cannot be chosen independently of the differences of the squared neutrino masses $\Delta {m}^2_{ij}$ due to the fact that the neutrino energy couples to the environment, whereas in our understanding this is not the case in \cite{Benatti:2000ph} demonstrating again that physical properties of the existing models crucially depend on the coupling to the environment. An interesting question is whether the final renormalised model in \cite{Benatti:2000ph} still allows massless neutrino oscillations.
~\\ ~\\
To also analyse to what extent the final model, and thus the counter term, depends on the choice of regulator, we considered four different choices, including the two most prominent ones, the Lorentz-Drude and the exponential cutoff of the Ohmic spectral density. As our results show, neither the explicit decoherence contribution nor the form of the counter term depend on this choice.
~\\ ~\\
One of the insights we have gained by taking a microscopic model as a starting point is that the physical interpretation of the individual contributions in the damping term that causes decoherence becomes clearer. Firstly, the model considered here contains only two free parameters, namely the coupling strength to the environment and the temperature of the thermal gravitational waves. If we assume that the latter can be obtained from independent experiments, then we are only left with the coupling strength between the neutrinos and the environment that enters the interaction Hamiltonian. Furthermore, the underlying gravitational coupling of matter to gravity that strongly inspired the toy model in  \cite{Xu:2020lhc} has the consequence, that exponential damping is determined by the difference of the neutrino energies squared.  In a next step we used this insight to perform a comparison to existing phenomenological models where usually a finite number of decoherence parameters is considered for which upper bounds are determined.
\\ \\
As our results show, we obtain a physical interpretation of the decoherence parameters used in the phenomenological models which is related to the coupling between the neutrinos and the environment in the microscopic model. From a theoretical point of view, we therefore expect that the limits for such decoherence parameters can be interpreted with a better physical understanding of the underlying microscopic model. Our analysis shows that in the case of vacuum oscillations we can achieve exact agreement with a subclass of phenomenological models. These are those that involve three neutrino flavors and for which at least three of the decoherence parameters, usually denoted as $\gamma_{ij}$ with  $\Gamma_{ij}=\gamma_{ij}E^n$, with $n \in (-2,-1,0,1,2)$, do not vanish, fulfil the relation to $\Delta  m^2_{ij}$ shown in \eqref{eq:connectionOurGammaPhenoGamma} and the dependence of $\Gamma_{ij}$ on the mean neutrino energy is given by $E^{-2}$, which corresponds to the phenomenological models $n=-2$. The fact that the power $n=-2$ is favored is a consequence of the coupling between the neutrinos and the gravitational wave environment. 
Hence, any other choice of power in the  phenomenological models will be rather difficult to be linked to gravitationally induced decoherence from our perspective. This is in contrast to the decoherence models inspired by quantum gravity in \cite{Ellis:1995xd,Ellis:1996bz,Ellis:1997jw,Ellis:2000dy}, which suggest $n=2$. The main difference to the model considered here is that in \cite{Ellis:1995xd,Ellis:1996bz,Ellis:1997jw,Ellis:2000dy} the second power of the energy, but not the energy difference, is included in the decoherence term. Our results agree with the model presented in \cite{Gambini:2003pv}, in which the squared energy difference is also included in the decoherence damping term. This sounds somewhat contradictory at first, but is resolved if we consider the system as neutrinos. Then, due to the squared energy difference, the linear term in the neutrino energy is the same for all neutrinos and simply cancels out, and what remains is a term proportional to $\Delta m^4_{ij}/E^2$. This shows that the power $n$ that is ultimately obtained depends crucially on both the choice of coupling to the environment and the choice of system, because for other than neutrinos the power will generally change, even with the same coupling to the environment is chosen, and the final form of the decoherence term depends on the resulting energy difference for the system under consideration.
~\\ ~\\
Furthermore, from our investigation we found that we are not able to match that subclass of phenomenological models just mentioned with the model presented here if we consider neutrino oscillations in matter. The reason for this is that in many existing phenomenological models the decoherence parameters $\gamma_{ij}$ do not depend on matter effects and thus are chosen to be constant for each layer of the Earth. In contrast, in the model presented here the neutrinos couple with their energy to the gravitational environment and the eigenvalues of the neutrino change if matter effects are present, thus the $\gamma_{ij}$ cannot be chosen to be constant across the different Earth layers. From the figures and the discussion in Sec.~\ref{sec:Results} it becomes clear that not taking such matter effects into account can lead to deviations in the probabilities for the neutrino oscillations of up to 10\% for the energies investigated in this work, depending on the chosen parameters of the model. That the decoherence term in the oscillation probabilities should be different for vacuum and matter was also a conclusion drawn in \cite{Carpio:2017nui}. However, to find an exact match with their models is not straight forward because they consider a perturbative expansion and consider only subleading decoherence terms.
~\\ ~\\
 At the level of the Lindblad equation, we can distinguish the two classes of models by a different choice of Lindblad operators. The subclass of phenomenological models is obtained by choosing the Lindblad operator as the vacuum Hamiltonian of the neutrino, while for the model presented here the Lindblad operator is the full Hamiltonian, which also contains matter contributions if they are present. Since we obtain these deviations in the probabilities for the neutrino oscillations, this is again an example of how different chosen couplings to the gravitational environment lead to different physical properties of the model. From this point of view, it is also clear why the models can match exactly in the case of neutrino oscillations in vacuum. From the standpoint of general relativity and linearised gravity respectively, a coupling with the vacuum Hamiltonian if matter is present seems not too obvious.
~\\ ~\\
Another consequence of this is that for the model presented here, the already existing bounds on the decoherence parameters $\gamma_{ij}$ could potentially be used to constrain the coupling strength with the gravitational environment (denoted as $\eta$ in our work). Indeed we have established the first experimental constraints based on the upper limits of the phenomenological model using KamLAND data \cite{DeRomeri:2023dht}
\\ 
We can further exploit matter effects to clearly distinguish between the phenomenological model and the model considered here in order to put direct bounds on our parameters. The effect appears to be visible with atmospheric neutrinos in the GeV energy range, propagating through the Earth. This makes neutrino telescopes such as KM3NeT/ORCA very good candidates to perform such a study. In this respect, we suggest such an analysis in order to determine the bounds on $\eta$ and $T$ for the model considered here.
~\\
~\\
Since we have mentioned that we understand the analysis of the toy model of \cite{Xu:2020lhc} in the context of neutrino oscillations as a first step to learn more about gravitationally induced decoherence, we would like to make some remarks on the limitations of our analysis and possible future steps. First, although the toy model is inspired by the field theoretical models in \cite{Anastopoulos:2013zya,Fahn:2022zql,Lagouvardos:2020laf,Oniga:2015lro,Blencowe:2012mp}, it is a quantum mechanical model. Starting with the field theoretical model and then projecting onto its 1-particle sector is more complicated and could generally lead to a more complex model that might contain additional features that we lose if we start directly with a quantum mechanical model. Hence, a careful analysis of the 1-particle sector of the model from \cite{Fahn:2022zql}, as done in \cite{Fahn:2024fgc} or see also \cite{Anastopoulos:2013zya,Lagouvardos:2020laf}, and a discussion of the results from \cite{Fahn:2024fgc} in the context of neutrino oscillations will therefore be part of our future work. Moreover, this will also allow to consider the renormalisation procedure in a broader context since it is quite simple in a quantum mechanical setup compared to quantum field theory. Moreover, one can analyse in parallel the different assumptions one makes, such as the first and second Markov approximation, to end up with the final Lindblad equation and learn whether the application of renormalisation before or after gives the same final one-particle sector. Finally, in order to generalise the model considered in this work, we also plan to generalise the quantisation procedure of the toy model so that we can also cover loop quantum gravity inspired models, which we study in the context of neutrino oscillations and compare with existing analyses such as the one in \cite{Al-Nasrallah:2021zie,DEsposito:2023psn}, where in \cite{Al-Nasrallah:2021zie} a minimum length model is analysed and \cite{DEsposito:2023psn} the decoherence effects are studied for different approaches in addition to \cite{Xu:2020lhc}, such as deformed symmetries \cite{Arzano:2022nlo}, metric fluctuations \cite{Breuer:2008rh} and fluctuating minimum lengths \cite{Petruzziello:2020wkd}.

\section*{Acknowledgements}
The authors would like to thank Th\'eophile Cartraud and Renata Ferrero for their contributions to our discussions in the initial and final stages of the project, respectively. The authors would also like to thank Renata Ferrero for her useful and valuable comments and suggestions for improvement on an earlier version of the manuscript. The authors also thank Stefan Hofmann for fruitful discussions on thermal gravitational waves in the cosmological context. A.~Domi would like to thank J.A.B.~Coelho for useful discussions on quantum decoherence in the neutrino oscillation context. This project has received funding from the European Union's Horizon 2021 research and innovation programme under the Marie Skłodowska-Curie grant agreement No. $101068013$ (QGRANT), supporting A.~Domi. M.J.~Fahn and M.~Kobler both thank the Heinrich-Böll foundation for financial support. 

\begin{appendices}

\section{Spectral Density from Field-theoretic model}\label{secA1}

To motivate the above choice of spectral density, in particular the linear dependence on $\omega$ for small $\omega$, and obtain a possible choice for $\eta$, we compare our toy model with the field-theoretic one in \cite{Fahn:2022zql}, where for simplicity we drop all indices in the latter. In the field-theoretic case, the role of the configuration variable is taken over by the densitised triad $\widehat{\delta E}$ and the canonically conjugated momentum is $\widehat{\delta C}$, which is a combination of the connection and the densitised triad. Their quantisation is introduced in \cite{Fahn:2022zql} in equations (3.6) and (3.7). To obtain the same commutation relations and environmental Hamiltonian as for the toy model discussed in this paper, we redefine $\widehat{\widetilde{\delta E}} := \sqrt{\kappa}\widehat{\delta E}$ and $\widehat{\widetilde{\delta C}} := \sqrt{\kappa}\widehat{\delta C}$, where $\kappa =\frac{8\pi G_N}{c^4}$ is the coupling constant in general relativity, containing Newton's gravitational constant $G_N$ and the speed of light $c$. The reason for this redefinition is the fact that their original algebra (see \cite{Fahn:2022zql} equation (3.3)) contains a factor $\kappa^{-1}$. In terms of these rescaled variables we obtain (see also \cite{Fahn:2022zql} equations (3.3) and (3.2)):
\begin{equation}
    \left[\widehat{\widetilde{\delta E}}, \widehat{\widetilde{\delta C}} \right] = i\hbar \delta \hspace{1in}  \hat{H}_E = \frac{1}{2}\int d^3k \left[ \widehat{\widetilde{\delta C}}(\vec k)^2 + \Omega_k^2 \widehat{\widetilde{\delta E}}(\vec k)^2 \right]\,,
\end{equation}
where the frequencies are defined as $\Omega_k := \sqrt{\vec k^2}$. In the interaction Hamiltonian (see \cite{Fahn:2022zql} equation (2.39) and below), one then couples the energy momentum tensor to the metric, which corresponds to contractions of $\widehat{\widetilde{\delta E}}$, hence
\begin{equation}
    \hat{H}_I \sim \int d^3k \frac{\sqrt{\kappa}}{2} \hat{T} \otimes \widehat{\widetilde{\delta E}}\,,
\end{equation}
where $\sim$ means "corresponds to". From this analogue one can deduce $g_i \sim \sqrt{\kappa}$. When computing correlation functions and hence $\Lambda$ and $\Gamma$, the terms appearing are of the form (see \cite{Fahn:2022zql} equations (4.38) and (3.6))
\begin{equation}\label{eq:appint}
    \int d^3k \frac{\kappa}{2\Omega_k} h(\vec k)
\end{equation}
compared to the expressions appearing here in \eqref{eq:lambdadef} and \eqref{eq:gammadef} using \eqref{eq:jsumdef}:
\begin{equation}
    \hbar^2 \sum_{i=1}^N \frac{ g_i^2}{2 \omega_i} h(\omega_i)\,.
\end{equation}
Assuming for simplicity that $h(\vec k) = h(\Omega_k)$, we can rewrite the integration in \eqref{eq:appint} in spherical coordinates:
\begin{equation}
    \int d^3k \frac{\kappa}{2\Omega_k} h(\Omega_k) = \int_0^\infty d\Omega\; 2\pi\kappa \Omega h(\Omega)\,.
\end{equation}
Motivated by this analogue, we take the following continuum limit for the toy model:
\begin{equation}
   \hbar^2  \sum_{i=1}^N \frac{g_i^2}{2 \omega_i} h(\omega_i) \longrightarrow \hbar^2 \int_0^\infty d\omega\; 2\pi\kappa \omega h(\omega)\,.
\end{equation}
This suggests to use as spectral density $J(\omega)$, which appears in integrals as $\int d\omega J(\omega) f(\omega)$, the following smooth function:
\begin{equation}
    J(\omega) = \hbar^2 \sum_{i=1}^N \frac{g_i^2}{2 \omega_i} \delta(\omega-\omega_i) \longrightarrow \frac{2\pi\kappa \hbar}{c} \omega\,,
\end{equation}
where we inserted additional inverse factors of $\hbar$ and $c$ to obtain the correct dimensions, as the work in \cite{Fahn:2022zql} is in natural units. To cure UV-divergence, we have to add a suitable cutoff, see the discussion in the main text. By comparison to the general form of the Ohmic spectral density in the main text, we find a possible choice for $\eta$ motivated by the analogy to the field-theoretic case:
\begin{equation}
    \eta^2 \sim \pi^2 \frac{\kappa\hbar}{2c} = 4\pi^3 \frac{\ell_p^2}{c^2} \approx 4\cdot 10^{-85}\,\mathrm{s^2}\,,
\end{equation}
where $\ell_p$ denotes the Planck length, thus
\begin{equation}
    \frac{\eta^2}{\hbar^2} \approx 10^{-54}\frac{1}{\mathrm{eV^2}} =  10^{-36}\frac{1}{\mathrm{GeV^2}}\,.
\end{equation}

\section{A note on the diagonalisation}\label{secADiag}
Due to the different orders of magnitude of $E$ compared to $\frac{\Delta m_{ij}^2 c^4}{6 E}$, working with $\hat{H}_{vac}$ as specified in \eqref{eq:hvaco} requires very high numerical precision and hence high computational effort. To simplify the computation, one can proceed in the following way: Splitting $\hat{H}_{S}^{(0)}$ into two parts in the following way:
\begin{equation}
    \hat{H}_{S}^{(0)} = E \mathds{1}_3 + \hat{H}_{S}^{(\Delta)}\,,
\end{equation}
the characteristic polynomial reads:
\begin{equation}
    \det \left( \hat{H}_{S}^{(0)} - \lambda \mathds{1}_3\right) = \det \left(\hat{H}_{S}^{(\Delta)} - \lambda^{(\Delta)} \mathds{1}_3 \right) = 0\,,
\end{equation}
where $\lambda^{(\Delta)} = \lambda - E$ are the eigenvalues of $\hat{H}_{S}^{(\Delta)}$. For the eigenvectors $v^{(\Delta)}$ of $\hat{H}_{S}^{(\Delta)}$ we find:
\begin{equation}
    \hat{H}_{S}^{(\Delta)} v^{(\Delta)} = \lambda^{(\Delta)} v^{(\Delta)} \iff \hat{H}_{S}^{(0)} v^{(\Delta)} - E v^{(\Delta)} = \lambda v^{(\Delta)} - E v^{(\Delta)} \iff \hat{H}_{S}^{(0)} v^{(\Delta)} = \lambda v^{(\Delta)}\,,
\end{equation}
thus $\hat{H}_{S}^{(0)}$ and $\hat{H}_{S}^{(\Delta)}$ have the same eigenvectors. Hence, one can work with $\hat{H}_{S}^{(\Delta)}$ and its eigenvectors throughout the calculation and just has to use in the final results in for instance \eqref{eq:solmeqeffmb} that $\widetilde{H}_i = \lambda_{{S},i}^{(\Delta)} + E$, where $\lambda_{{S},i}^{(\Delta)}$ are the eigenvalues of $\hat{H}_{S}^{(\Delta)}$. As the evolution of the neutrino in the end only depends on terms of the form $\left(\widetilde{H}_i-\widetilde{H}_j\right)$, the mean neutrino energy $E$ will always cancel.

\section{Solution of the Master Equation}\label{secA2}

The master equation in effective mass basis ~\eqref{eq:masterequeffmb} is
\begin{equation}
    \frac{d}{dt}\rho(t) =-\frac{i}{\hbar} \left[ H, \rho(t)\right] +l \left(H \rho(t) H -\frac{1}{2} \left\{ H^2, \rho(t) \right\} \right)
\end{equation}
with the scalar prefactor $l = \frac{8\eta^2}{\hbar^2} \frac{k_B T}{\hbar}$. For better readability, we dropped all hats and tildes and indices, also for $H:= \hat{\widetilde{H}}_S$. To solve this equation, we consider the three summands individually. As all operators that appear apart from the density matrix commute and are time-independent, we can solve the master equation for each summand individually and then combine the solutions.
\begin{itemize}
    \item For the first summand we find:
    \begin{equation}
        \frac{d}{dt}\rho^{(1)}(t) =-\frac{i}{\hbar} \left[ H, \rho^{(1)}(t)\right] \hspace{0.1in} \implies \hspace{0.1in} \rho^{(1)}(t) = e^{-\frac{i}{\hbar} H t} \rho^{(1)}(0) e^{\frac{i}{\hbar} H t}\,.
    \end{equation}
    \item The third summand yields:
    \begin{equation}
        \frac{d}{dt} \rho^{(2)}(t) =- \frac{l}{2} \left\{H^2, \rho^{(2)}(t)\right\} \hspace{0.1in} \implies \hspace{0.1in} \rho^{(2)}(t) = e^{- \frac{l}{2}H^2 t} \rho^{(2)}(0) e^{-\frac{l}{2} H^2 t}\,.
    \end{equation}
    \item For the second summand we obtain:
    \begin{align}
        \frac{d}{dt}\rho^{(3)}(t) = l \cdot  H \rho^{(3)}(t) H \hspace{0.1in} \implies \hspace{0.1in}  \rho^{(3)}(t) = &\sum_{n=0}^\infty \frac{(l t)^n}{n!}  H^n \rho^{(3)}(0) H^n 
    \end{align}
\end{itemize}
Combining these in a suitable form, the total solution then reads:
\begin{align}
    \rho(t) &= \sum_{n=0}^\infty \frac{(l t)^n}{n!} H^n  e^{-\frac{i}{\hbar} H t -\frac{l}{2} H^2 t} \rho(0) e^{\frac{i}{\hbar} H t -\frac{l}{2} H^2 t}   H^n \nonumber\\
    &=: \sum_{n=0}^\infty \frac{(l t)^n}{n!}  H^n  F \rho(0) F^\dagger H^n \,,
\end{align}
where we defined $F :=  e^{-\frac{i}{\hbar} H t -\frac{l}{2} H^2 t} $. Since $F$ and $H^n$ are diagonal in the effective mass basis, we can evaluate the matrix product directly and obtain:
\begin{equation}
    \rho_{ij}(t) = \sum_{n=0}^N \frac{(l t)^n}{n!} \rho_{ij}(0) F_i F_j^* H_i^n H_j^n = \rho_{ij}(0) F_i F_j^*  e^{l H_i H_j t}\,,
\end{equation}
where a star denotes complex conjugation and we refer to the components of the diagonal matrices as $F_i := F_{ii}$ and $H_i := H_{ii}$. Reinserting the original expressions, this yields the solution \eqref{eq:solmeqeffmb}:
\begin{equation}
\widetilde{\rho}_{ij}(t) =\widetilde{\rho}_{ij}(0) \cdot e^{ -\frac{i}{\hbar} \left( \widetilde{H}_i - \widetilde{H}_j \right) t - \frac{4\eta^2}{\hbar^2} \frac{k_B T}{\hbar} \left( \widetilde{H}_i -\widetilde{H}_j \right)^2 t } \,.
\end{equation}

\end{appendices}

\bibliography{sn-bibliography}% 

\end{document}